\documentclass[a4paper,fleqn,usenatbib]{mnras}
\usepackage[T1]{fontenc}
\usepackage{ae,aecompl}
\usepackage{graphicx}    
\usepackage{amsmath}    
\usepackage{amssymb}    

\usepackage{longtable}
\usepackage[para]{threeparttable}
\usepackage[normalem]{ulem}
\usepackage{lmodern}

\usepackage{color}

\newcommand{\hii}{\mbox{H~{\sc ii}~}}

\title[Variable Stars in Cyg 0B7]{Optical Photometric Variable Stars towards  Cygnus OB7}
\author[S. Dutta et al.]{Somnath Dutta$^{1,2}$\thanks{E-mail: duttasomnath9@gmail.com (SD)},  Soumen Mondal$^{1}$,  Santosh Joshi$^{3}$ and Ramkrishna Das$^{1}$\\
$^{1}$S.N. Bose National Centre for Basic Sciences, Kolkata 700 106, India\\
$^{2}$Academia Sinica Institute of Astronomy and Astrophysics, P.O. Box 23-141, Taipei 106, Taiwan\\
$^{3}$Aryabhatta Research Institute of Observational Sciences, 
Nainital-263 002, India}

\usepackage{graphicx}
\begin{document}



\maketitle


\begin{abstract}
 We present optical $I$-band light curves of the stars towards a star-forming region Cygnus OB7 from 17 nights photometric observations. The light curves are generated from a total of 381 image frames with very good photometric precision. From the light curves of 1900 stars and their periodogram analyses, we detect 31 candidate variables including five previously identified. 14 out of 31  objects are periodic and exhibit the rotation rates in the range of 0.15 to 11.60 days. We characterize those candidate variables using optical/infrared colour-colour and colour-magnitude diagrams. From spectral indices of the candidate variables, it turns out that four are probably Classical T-Tauri stars (CTTSs), rest remain unclassified from present data, they are possibly field stars or diskless pre-main sequence stars towards the region. Based on their location on the various colour-magnitude diagrams, the ages of two T Tauri Stars were estimated to be  $\sim$ 5 Myr. The light curves indicate at least five of the periodic variables are eclipsing systems. The spatial distribution of young variable candidates on Planck 857 GHz (350 $\mu$m) and 2MASS $K_s$ images suggest that at least two of the CTTSs are part of the active star-forming cloud Lynds 1003.

\end{abstract}


\section{Introduction} 

\begin{figure*}
\includegraphics[width=19cm,height=10.0cm]{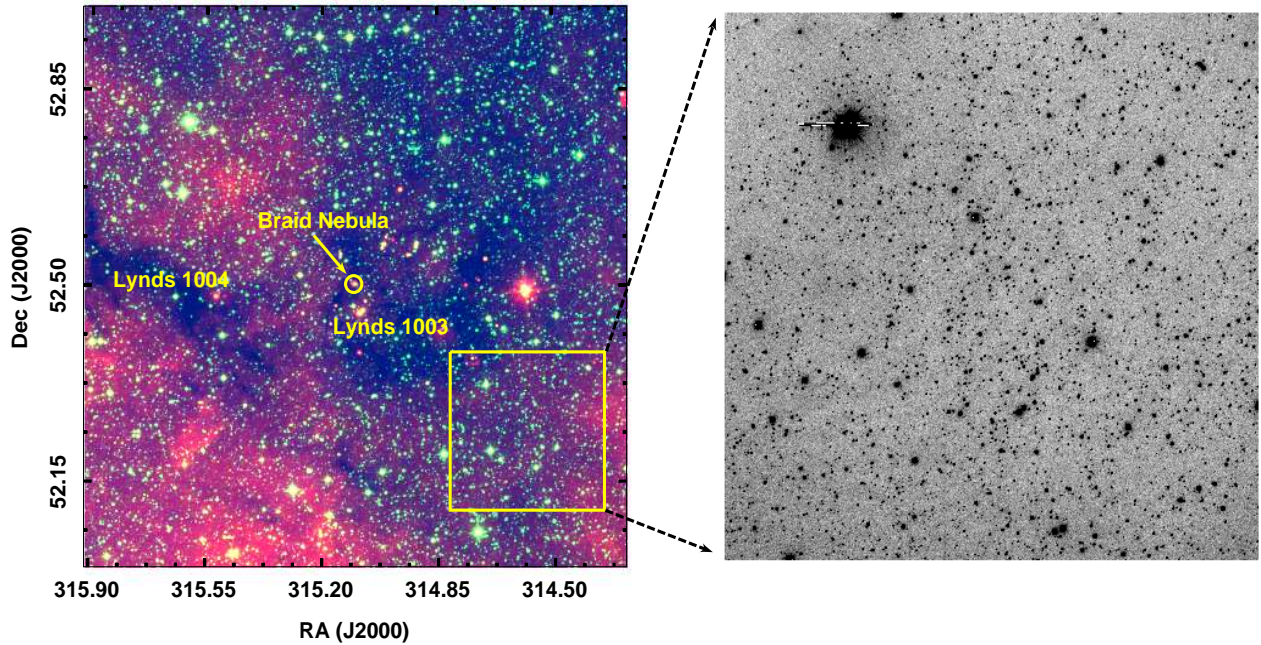}
 \caption{({\it left panel}) Infrared view (red: WISE W3, green: WISE W1, blue: 2MASS $Ks$) of dark cloud Lynds 1003 and Lynds 1004 towards Cyg OB7 region. The braid nebula is also marked. Our optical FOV is marked in yellow square. ({\it right panel}) Optical $I$-band image of our studied region (yellow square in the left panel) was observed using 1.3m DFOT. See text for details.}  
  \label{fig:observed_field_spatial}  
\end{figure*}

Pre-main sequence (PMS) stars first came into the spotlight due to their photometric variability characteristics \citep{1945ApJ...102..168J, 1962AdA&A...1...47H}. Since then numerous observational studies have been performed on exploring variability in young stars and the role of angular momentum in their stellar evolution \citep[e.g.,][]{2000AJ....120..349H,2001AJ....121.3160C,2004A&A...417..557L,2004AJ....127.2228M,2005MNRAS.358..341L,2005AJ....129..907B,
2017ApJ...836...41C}. The variation in the observed flux of a PMS star is thought to be originated via various mechanisms such as magnetically induced cool starspots or magnetically channelled variable accretion flows generating hot spots on the star surface, eclipsing binary (EB), opacity due to non-uniform dust distribution, etc. \citep{1999AJ....117.2941S, 2002A&A...396..513H}.  Each of these crucial components encounters various types of dynamic phenomena at a different wavelength and may induce variations in observed flux. In general, Classical T-Tauri Stars (CTTSs) having stronger emission lines (EW $>$ 10 \AA) \& high infrared excess show irregular variability, whereas Weak-line T-Tauri Stars (WTTSs) having  weak H$\alpha$ emission (EW $<$ 10 \AA) \& small/no infrared excess are mostly periodic. However, several systematic studies on PMS star reveal that  the CTTS could also exhibit periodic variability  
\citep[e.g.,][]{1983ApJ...266L..45S,1993A&A...272..176B,2006PASP..118.1390P,2010PASP..122..753P,2018MNRAS.476.2813D}

The Cygnus rift consists of several young star-forming regions \citep[SFRs;][]{2008hsf1.book...36R}. The Cygnus OB7 (Cyg OB7) is the nearest of all recognized Cygnus OB associations, which is located at $\sim$ 760 pc  \citep{1956ApJS....2..389H, 1958AN....284...76S}.   A number of young embedded sources (e.g., CTTSs, Herbig Ae/Be) were identified in the cloud complexes of Cygnus \citep[e.g.,][]{1980AJ.....85...29C,1988cels.book.....H,1997IAUS..182P..91D,2003Ap.....46..282M,2006A&A...455.1001M,2013MNRAS.432.2685M,2016Ap.....59..484M}. The region harbours many  IRAS sources, young massive OB stars including outflowing candidates, and thick disk stars. The Cyg OB7 region contains numerous dark cloud complexes \citep[][]{1962ApJS....7....1L}. The dark cloud Lynds 1003, located at northern part of Cyg OB7, was first investigated by \citet[][]{1980AJ.....85...29C} and later by several authors \citep[][]{2001Ap.....44..216M,2003Ap.....46..282M,2008arXiv0810.3943A,2013ApJ...773..145W}. Several young stellar objects (YSOs) were identified towards the region. \citet[][]{2006A&A...455.1001M} presented the evidence of a young star illuminating the Braid Nebula (see {\it left panel} of Figure~\ref{fig:observed_field_spatial} ), which is possibly an eruptive variable of the FU Orionis (FUor) type. The near-Infrared (NIR) $JHK$ monitoring of the dark clouds Lynds 1003/1004 in the Cyg OB7 were carried out by \cite{2012ApJ...755...65R}, \cite{2013AJ....145..113W} and \cite{2013ApJ...773..145W} in search for disk-bearing young stars as well as variable stars in an approximate region of   Figure \ref{fig:observed_field_spatial} ({\it left panel}). In this paper, we performed optical $I$-band monitoring observations to detect variability of young stars in the outer part of Cyg OB7 regions since the region is relatively less extinct in optical bands and large populations of young stars could be attempted.

 Identification of variable candidates needs long-term monitoring over a large number of stars. 
Such investigation involved not only for characterizing variability but also for detecting interesting and rare objects with their rare phenomena. The principal focus of the  SFR monitoring is related to many questions as the fraction of young stars that exhibit variability, the amplitude and rotation period of photometric fluctuation, and dominant physical mechanisms related to it. In this contribution, we present  here optical $I$ time-series photometry over an area $\sim$ $15\arcmin \times 15\arcmin$ centred on $\alpha_{2000}$ = $20^h58^m28^s$ $\delta_{2000}$ = $+52^015^m26^s$) towards south-western outskirts of Lynds 1003 cloud in Cyg OB7, overlapping with some area monitored by \cite{2012ApJ...755...65R} and \cite{2013ApJ...773..145W} in the NIR. The field-of-view (FOV) of our studied region is displayed in the {\it right panel} of Figure \ref{fig:observed_field_spatial}. Section \ref{sec:data} describes the data set obtained for this study and the data reduction procedure. Section \ref{sec:variable_iden} deals with the identification of variable stars from their time-series photometry and the different types of the observed variability. In Section \ref{sec:discussion}, we discuss the characteristics of the candidate variables including the possible correlation between variability and circumstellar disk material, and the spatial distribution of variable stars in the molecular clouds. Our conclusions are summarized in section \ref{sec:summary}.

\begin{table*}
\centering
 \begin{minipage}{140mm}
  \caption{Log of Observations.} 
\label{tab:observation} 
\begin{tabular}{|c|c|c|c|c|c|}
\hline

 Date of      & Telescope &  I                   & R        &  V   & Avg. seeing \\
 Observations &           & Exp.(s)$\times$ N &Exp.(s)$\times$ N & Exp.(s)$\times$ N &    (arccsec)\\
\hline
\hline

31.03.2014&1.3m DFOT&120$\times$22, 60$\times$1 & 250$\times$2, 150$\times$1& 300$\times$1, 100$\times$1&  1.8\\
13.11.2014&1.3m DFOT&150$\times$42, 90$\times$20, 60$\times$25 & $...$&$...$ &  1.8\\
14.11.2014&1.3m DFOT&150$\times$15, 90$\times$1, 60$\times$8  &$...$ &$...$ &  1.8\\
29.11.2014&1.3m DFOT&150$\times$4, 90$\times$3, 30$\times$3 &$...$ & $...$&  2.2\\
10.12.2014&1.3m DFOT&150$\times$13, 90$\times$10, 20$\times$5 & $...$&$...$ &  1.8\\
11.12.2014&1.3m DFOT&150$\times$7, 60$\times$5, 20$\times$6& $...$ & $...$&  1.8\\
\hline
21.05.2014&2.0m HCT&120$\times$21  & $...$&$...$ & 2.2\\
19.08.2014&2.0m HCT&120$\times$10, 90$\times$6, 60$\times$8  & $...$&$...$ & 1.0\\
29.10.2014&2.0m HCT&120$\times$3, 90$\times$3, 60$\times$5  & $...$&$...$ & 2.5\\
30.10.2014&2.0m HCT&120$\times$5, 60$\times$3  & $...$&$...$ & 2.5\\
05.10.2015&2.0m HCT&300$\times$2, 150$\times$4  &300$\times$1, 60$\times$1, 10$\times$1&$...$ & 2.5\\
06.10.2015&2.0m HCT&300$\times$2, 150$\times$4, 60$\times$4  &300$\times$1, 30$\times$1 &$...$ & 2.5\\
07.10.2015&2.0m HCT&300$\times$2, 150$\times$3, 60$\times$2  &300$\times$1, 30$\times$2  & $...$& 2.5\\
\hline
28.05.2014&1.04m ST&150$\times$12, 90$\times$3, 30$\times$2&$...$ &$...$ & 2.5\\
29.05.2014&1.04m ST&150$\times$26                           &$...$ &$...$ &2.5\\
03.06.2014&1.04m ST&150$\times$18, 30$\times$3& 200$\times$1, 90$\times$1 & 250$\times$1 & 2.5\\
05.06.2014&1.04m ST&200$\times$5, 150$\times$31, 60$\times$4 &$...$ &$...$  & 2.5\\

\hline
\hline\end{tabular}
\end{minipage}
\end{table*}

\section[]{Data sets}
\label{sec:data}
\subsection{Observations and data processing}
\label{sec:optical} 
 Photometric $V$, $R$ and $I$ observations were carried out on 17 nights spanning over 1.5 years during 2014-2015 with three different telescopes. The $I$ band data were used for monitoring variability.

We performed observations using the 2K $\times$ 2K CCD camera on the 1.3 m Devasthal Fast Optical Telescope (DFOT), ARIES, Nainital, India. The FOV of the camera was $\sim$ 18$\arcmin \times 18\arcmin$ with a plate scale of 0.535 arcsec pixel$^{-1}$. More detailed on instrument specification and observations technique are described in \citet[][]{2018MNRAS.476.2813D}. We carried out further {\it I}-band monitoring observations using the 2K $\times$ 2K CCD camera of the HFOSC instrument at the 2 m Himalayan Chandra Telescope (HCT), India, which has an FOV $\sim$ 10$\arcmin \times 10\arcmin$ with a pixel scale of 0.296 arcsec pixel$^{-1}$ \citep[see][for details]{2018MNRAS.476.2813D}. We observed another set of {\it I}-band monitoring observations at the 1.04 m Sampurnanand Telescope (ST) operated by ARIES. The 2K $\times$ 2K CCD camera has an FOV about 13$\arcmin \times 13\arcmin$ with a plate scale of 0.37 arcsec pixel$^{-1}$ \cite[see][for more details]{2015MNRAS.454.3597D}.

All the targets and comparison stars were chosen within the linear response curve of the CCD camera for our differential (relative) photometry, for which the observations were taken in both long as well as short exposures. Thus, we achieved good dynamic coverage of the bright as well as faint sources. We always prefer dark to grey nights for our observations. The sky was clear in most cases and the target field was always far from the Moon during observing nights.
The 2$\times$2 binning mode was applied during observations to get a better signal-to-noise ratio (SNR). The typical FWHM of the stars were in the range $1.0-2.5 \arcsec$ during our whole observing span. We provide the log of observations in Table~\ref{tab:observation}, which includes the date of observations at different telescopes; the number of frames with different exposures in different filters; the average seeing at each night. 

The prepossessing of raw images were performed with standard reduction steps in IRAF\footnote{Image Reduction and Analysis Facility (IRAF) is distributed by National Optical Astronomy Observatories (NOAO), USA (http://iraf.noao.edu/)} software. The image frames were cleaned following bias subtraction, flat-fielding and cosmic ray removal using default tasks available in {\it noao.imred}. The point sources were identified with the {\it daofind} task of DAOPHOT package. The photometry was calculated by point source function (PSF) fitting using {\it allstar} task of DAOPHOT package \cite[][]{1992ASPC...25..297S}.

\begin{figure}
\includegraphics[width=8.0 cm,height=8.5cm, angle=0]{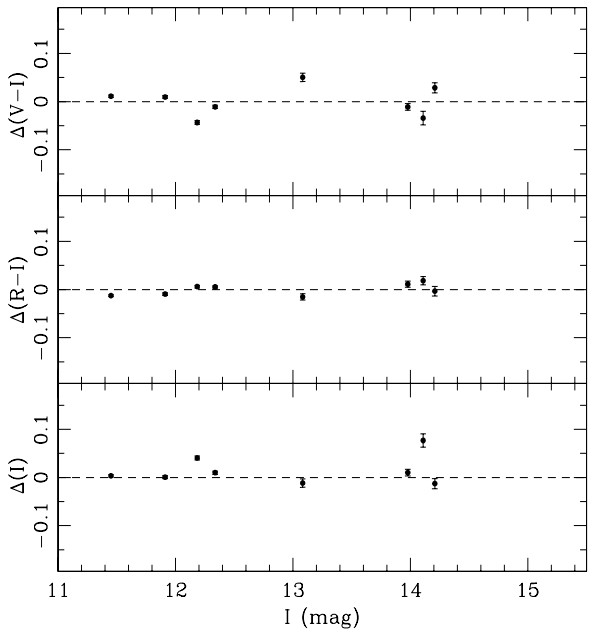}
  \caption{The standardization residuals between transformed and standard magnitudes and colours of observed standard stars are displayed. The combined error bars were propagated from the errors of transformed magnitudes and Landolt (1992).}  
  \label{fig:residual}  
\end{figure}

The instrumental magnitudes were transformed into standard magnitudes following the procedure described in \citet[][]{1987PASP...99..191S}. We observed  a set of 8 optical standard stars in RU 149 field \cite[][]{1992AJ....104..340L} using DFOT\footnote{We transformed only DFOT instrumental magnitudes to standard magnitudes, since standard magnitudes are utilized for optical colour-magnitude diagram only and the DFOT has largest FOV.} to estimate the atmospheric extinction and transformation coefficients. The photometric calibrations were performed for the DFOT observations by the following transformation equations:

\begin{equation}
(V-I) = (0.781 \pm 0.007)(v-i) + (0.170\pm0.006)
\end{equation}
\begin{equation}
(R-I) = (0.650\pm 0.008)(r-i)  + (0.455 \pm 0.010)
\end{equation}
\begin{equation}
I = i + (0.975 \pm 0.001)(r-i) + (-2.410\pm 0.002) 
\end{equation}
where $V$, $R$ and $I$ are the standard magnitudes and $v$, $r$, $i$ are the instrumental magnitudes corrected for the atmospheric extinction. The errors  in the final magnitudes were propagated from profile-fitting photometry, standard coefficients and uncertainties in the extinction coefficients.  
In Fig.~\ref{fig:residual}, the residuals ($\Delta$) of transformed and standard $I$ magnitudes, $(R - I)$ and $(V - I)$ colours of standard stars are shown as a function of $I$ magnitudes. The standard deviations in $\Delta I$, $\Delta (R - I)$ and $\Delta (V - I)$  are estimated to be 0.020, 0.018 and 0.020 mag, respectively.  

Finally, we estimated optical magnitudes of the 7009  objects detected in both  $R$ and $I$ bands. The $I$-band light curves of those detected stars were generated from 381 imaging frames using 17 nights photometric observations on different telescopes. Nevertheless, we considered 1900 stars for variability studies having at least 140 data points in the overlapping region of three telescopes (Table~\ref{tab:observation}). The completeness limits at various bands were estimated from the histogram turn-over method \cite[e.g.,][]{2015MNRAS.454.3597D}. Our photometric data are complete down to $V$=21 mag, $R$=21 mag, $I$=20.5 mag.

 The usage of multiple telescopes makes scaling an issue since the colour transformation varies from the telescope to telescope. So the differential photometry has an added systematic term (magnitude offset). Additionally the variable stars could have colour variation. Thus, a single set of transformation equation is not sufficient to convert time-series instrumental magnitudes into their standard form. So, we prefer instrument magnitudes for further estimation of variability characteristics. Such systematic variation between two different telescopes could be further reduced by the offset of magnitudes for a set secondary standard star (a list of non-variable stars; see also section \ref{sec:relative}) observed with those different telescopes.  We have considered DFOT magnitudes as primary (since the most number of observations) and applied the estimated offsets for the other two telescopes (HCT \& ST). However, if we till accept that there is a significant affect of the colour term in the investigation of variable stars, less than 5\% of our sample will have contaminated light curves. 
 

 The world coordinate system (wcs) coordinates of the observed stars were estimated using IRAF tasks $ccmap$ and $ccsetwcs$. We selected the positions of 25 isolated moderately bright stars from the 2MASS point source catalogue \cite[PSC;][]{2003yCat.2246....0C} to obtain the astrometric solution; which allow us to achieve very good positional accuracy ($< 0.5\arcsec$).

\subsection{Archival data set}

The NIR observations towards Cyg OB7 (centered on $\alpha_{2000}$ = $20^h58^m28^s$ $\delta_{2000}$ = $+52^015^m26.0^s$) were obtained  in $JHK$ bands from the UKIRT Infrared Deep Sky Survey (UKIDSS) data archive observed using WFCAM camera of the  3.8 m UKIRT telescope. In this survey telescope each pixel corresponds to 0.3$\arcsec$ and yields an FOV $\sim$ 20$\arcmin$ $\times$ 20$\arcmin$. The average FWHM during the observing period was $\sim$ 1.2$\arcsec$. The identification of point sources and photometry were performed using IRAF with the same procedure described in section \ref{sec:optical}.

  To avoid the inclusion of UKIRT saturated sources, we replaced all the sources in our catalog having 2MASS magnitudes less than $J$ = 13.25, $H$ = 12.75 and $K$ = 12.0 mag, respectively \cite[][]{2008MNRAS.391..136L}. The magnitude of the sources with uncertainty $\le$ 0.1 mag was taken for our study to ensure good photometric accuracy.

 The WISE survey provides photometry at four wavelengths 3.4 (W1), 4.6 (W2), 12 (W3) and 22 (W4) $\mu$m, with an angular resolution of 6.1\arcsec, 6.4\arcsec, 6.5\arcsec, 12.0\arcsec, respectively \cite[][]{2012yCat.2311....0C}.  We used WISE photometric catalog with good quality photometry (uncertainty better than 20\%) in this study.

  The Planck  satellite observed  the  entire sky in nine frequency wavebands during 2009-2013 \citep[][]{2011A&A...536A...1P,2016A&A...586A.134P} with two instruments: the High Frequency Instrument \citep[HFI; 857,  545,  353,  217,  143  and  100 GHz;][]{2010A&A...520A...9L} and the  Low  Frequency  Instrument.  For this analysis, we used the 857 GHz (350 $\mu$m) map.

\section{Analysis}
\label{sec:variable_iden}

\begin{figure}

\includegraphics[width=9 cm,height=6.5cm]{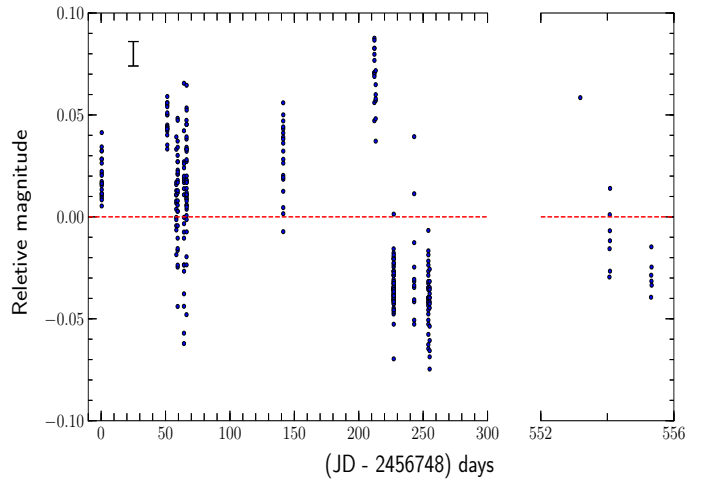}
  \caption{ The observed light curve  is shown  for the ID6828 of mean $I_{inst}$ = 18.93 mag ($I_{std}$ = 17.05 mag) with an RMS = 0.05 mag. The rms of all photometric precesion (= 0.015 mag) is also marked in top-left corner (see text for details).}  
  \label{fig:id6828_irregular}  
\end{figure}

\subsection{Relative photometry}
 \label{sec:relative}  
 
   Owing to the good quality of the light curves, we estimated relative photometry using the instrumental magnitudes, which reduced the effect of the sky background variation, airmass effect, signature of instrument etc. First, we searched a set of non-variable reference stars having stable photometric light curves with relatively less root-mean-square (RMS) deviation   and no periodicity on their frequency distribution (discussed later). The relative magnitudes ($\Delta m_i$) of each target star were estimated by subtracting target star magnitudes ($Mt_i$) from reference star magnitudes ($Mc_i$) for all the observed images (i = 1, 2, 3, .... N). Thus, final differential light curves were obtained by using the equation:

    \begin{equation}
     \Delta m_{i} = Mt_{i} - Mc_{i}, ~~ i = 1, 2, 3, .... N
    \end{equation}

 We selected here five such reference stars for our analysis to produce a set of five independent light curves for each target star; which helped us to cross-verify and rule out any possible contamination from reference stars.
An example light curve is shown in Figure~\ref{fig:id6828_irregular} to illustrate the cadence of the data, where we have estimated a mean instrumental magnitude of $I_{inst}$ = 18.93 mag with an RMS = 0.05 mag for the star ID\,6828. 
The errors within a night have an RMS of about 0.015 for ID6828; which would also vary from star to star for different magnitude range. The intra night precision is 18 mmag for this example light curves.  On the average, the intra-night precision vary as 5 mmag for $I_{inst}$ =15 mag; 20 mmag $I_{inst}$  = 19 mag; 70 mmag for $I_{inst}$  = 22.5 mag; thus we achieve very good photometric precision in our overall observed data points even in the high nebulous and crowded region. The mean intra night error of magnitudes is 0.019. Finally, after folding all 17 nights data, we estimated an RMS of magnitudes of about 50 mmag for the above example.
 

\subsection{Selection of variable candidates}
 \label{sec:scatter_in_mag} 
\begin{figure}
 \centering
 \includegraphics[width=8.5 cm,height=6.0cm]{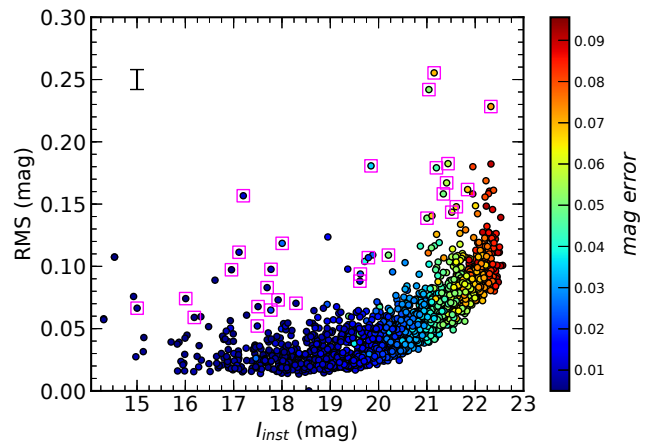}
  \caption{RMS distribution of the $I_{inst}$ magnitudes of 1900 stars towards Cyg OB7 is shown. The magenta boxes are the candidate variables identified from their observed light curves. An average value of error in RMS is plotted in the top left corner. The colour bar indicates the average magnitude error associated to each magnitude (see text for details).}
  \label{fig:rms}
\end{figure}

 The identification of variable stars was performed by means of careful inspection of the light curves. First, we estimated RMS ($\sigma_{mag}$) deviation of relative magnitudes for all the stars \citep[e.g.,][]{2018MNRAS.476.2813D}. In the Figure \ref{fig:rms}, the lower envelope represents an expected trend of $\sigma_{mag}$ as a function of stellar brightness, where $\sigma_{mag}$ increases with $I_{inst}$ magnitudes since SNR decreases accordingly.
  The $\sigma_{mag}$ values range from $\sim$0.013 mag for the bright stars to $\sim$0.12 mag for stars towards completeness limit (see section \ref{sec:optical}). Eventually, a few of them are scattered from the normal trend. These scattered outliers may arise from several facts e.g., intrinsic variability of the star itself, photometric noise, variation in sky signal. Moreover, visual inspection of observed light curves (section \ref{sec:relative}) indicated that a few stars show large RMS scatter because of abrupt flux change of only a few data points. Nonetheless, such spurious measurements were rejected as their false variability comes from their location on the edge of CCD, the effect of the bad pixel or cosmic ray hits.

\begin{figure}
 \includegraphics[width=8 cm,height=10.0cm]{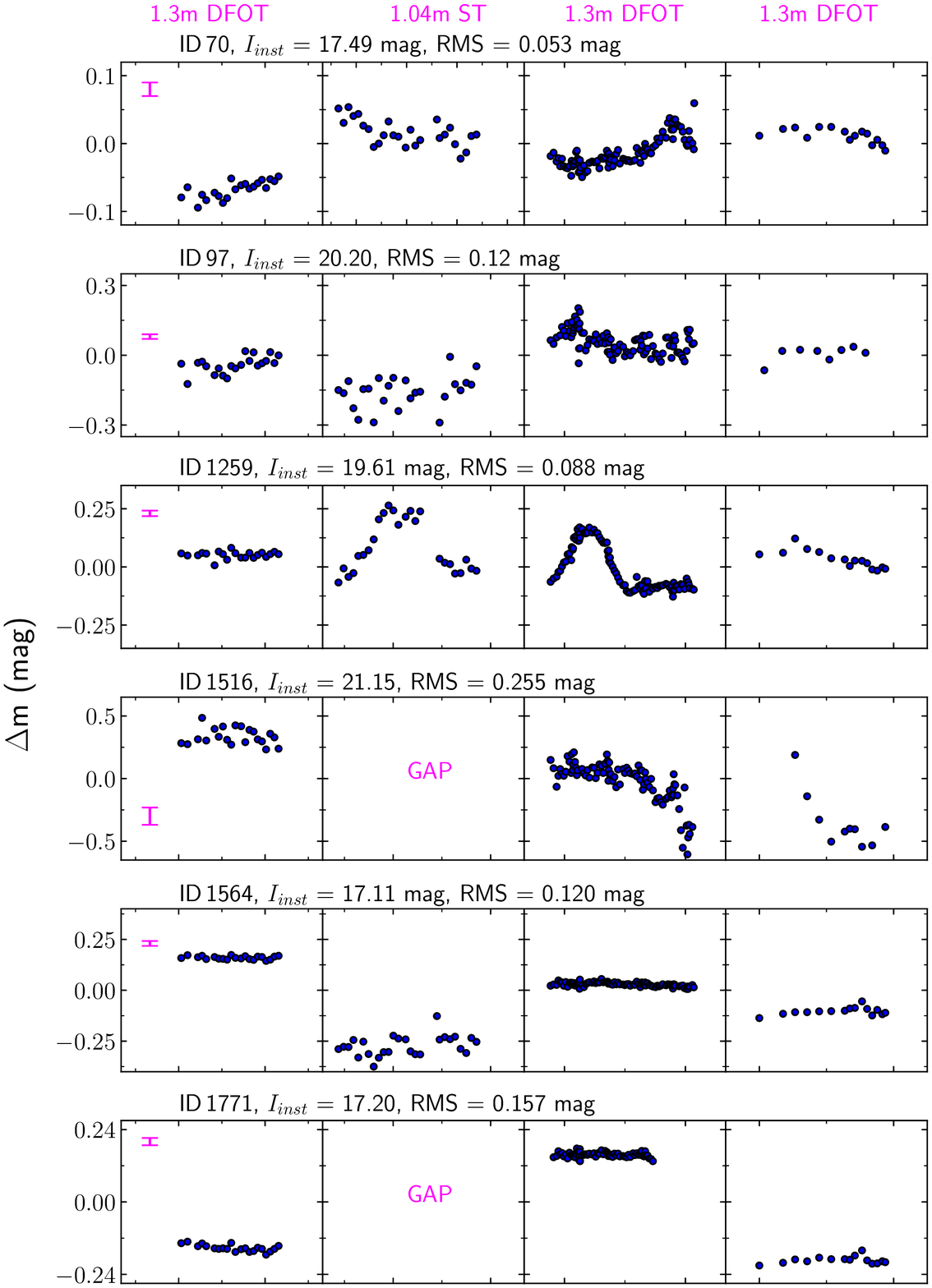}
 \includegraphics[width=8 cm,height=10.0cm]{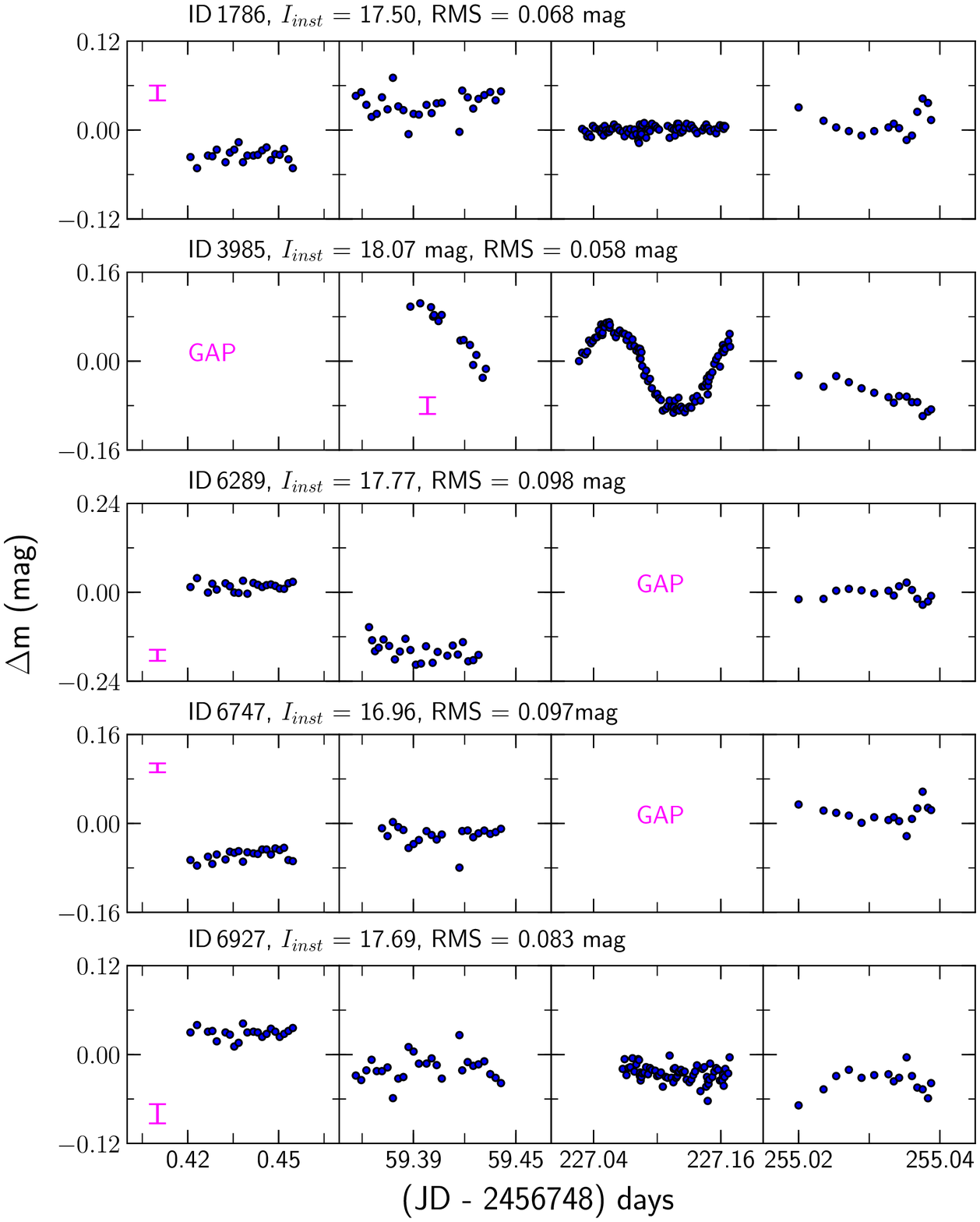}
  \caption{A few examples of differential light curves are plotted. The telescopes of each dataset are marked on top of each panel. The average error bar (in magenta) for each star is shown in the first panel for each row. The `GAP' indicates the absence of data points on that night (see text for details).}
  \label{fig:obs2_variable_interesting}
\end{figure}

 The candidate variables were selected from the observed 1900 light curves based on 3$\sigma$ RMS scatter from the mean in each magnitude bin (see Figure \ref{fig:rms}). From Figure \ref{fig:rms}, it is clear that all the candidate variables are lying above $\sim$ 0.01 mag from the ``banana''-shaped pattern of non-variable stars, however, the variables were further authenticated  with their observed light curves by visual inspection.
Finally, we identified a total of 31 variable stars from the light curve analyses. A few observed light curves of variable stars are shown in Figure \ref{fig:obs2_variable_interesting} for four-night observations, to reflect our confidence of visual inspection during variable selection. The average errorbars corresponding to each star is displayed. The `GAP'\footnote{Since, we have observed with different telescopes with different FOV and also at varying central position to cover all nearby objects in our studied region, a few of the stars may locate at the edge of the CCD, which we remove from further analyses. Moreover, our observations have been carried out with different exposures to achieve good coverage of the  bright as well as the faint sources, the total number of points of observations may vary to maintain a good photometric precision for all the data points.} represents the absence of data points on that night. The catalog of variable stars and their light curve analysis are presented in Table~\ref{tab:cat_var} and Table~\ref{tab:parameters}, respectively.

\begin{figure}
 \includegraphics[width=8 cm,height=6.0cm]{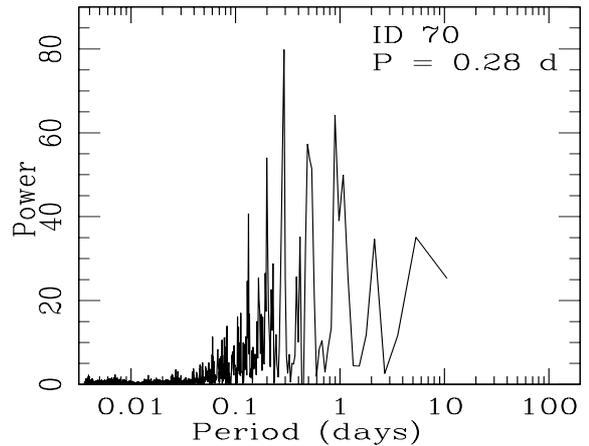}
  \caption{Example LS power spectra of a variable star. The highest peaks correspond to the period of the star.}
  \label{fig:freq}
\end{figure}

\subsection{Period Estimation}
\label{period_estimation}
   We searched for the periodic signals in all of the monitored stars independent of their observed light curve and $\sigma_{mag}$ deviation.
    We made use of the Lomb-Scargle (LS) periodogram analysis technique \citep[][]{1976Ap&SS..39..447L, 1982ApJ...263..835S}. The LS method is used to find out significant periodicity even with unevenly sampled data and was verified successfully in several cases to determine periods from such sparse data sets \citep[e.g.,][]{2004A&A...417..557L, 2010AJ....139.2026M,2012MNRAS.427.1449L}. This method calculates the normalized power for a given angular frequency and locates the highest peak in the estimated periodogram. This highest peak is considered as the most significant peak (False Alarm Probability < 0.01; FAP) and provides the period of the time-series data.

     The highest frequency corresponds to the Nyquist frequency, which is measured on the basis of  average observational spacing.
    The periodogram peak could fall in the gap between the chosen grid points of frequency.  So, a reliable approach is to use the frequency heuristic to decide on the appropriate grid spacing to use, or passing a minimum and maximum frequency. The minimum frequency corresponds to the time range and the maximum frequency is obtained from the median of consecutive observational spacing. 
Alternatively, to achieve a more precise estimate of the periods, we chose a very small step size of frequency to smoothly sample the periodogram as we zoom in towards the period. In this approach, the step size of the period search algorithm is a function of the square of the period. We consider a step size of $\sim$ 0.001 day for periods less than 3 days and $\sim$ 0.015 days for a period up to 12 days. 
The derived (linear) frequencies has the formal error, $\delta$f = 3$\sigma_N$/2TA $\sqrt{N_0}$ \citep[][]{1986ApJ...302..757H}, where $\sigma_N$ is the variance of noise in the period subtracted signal, T is the span of time baseline ($\sim$ 550 days), A is the signal amplitude (typically 10\%), $N_0$ is the number of independent data points ($\sim$ 300). Thus, we obtain the formal error of the order of $\sim$ 0.001 days. We measure the periods nearest to 0.01 days, which is quite reasonable for our period measurements of less than 12 days.

\begin{figure}
  \includegraphics[width=8 cm,height=10.0cm]{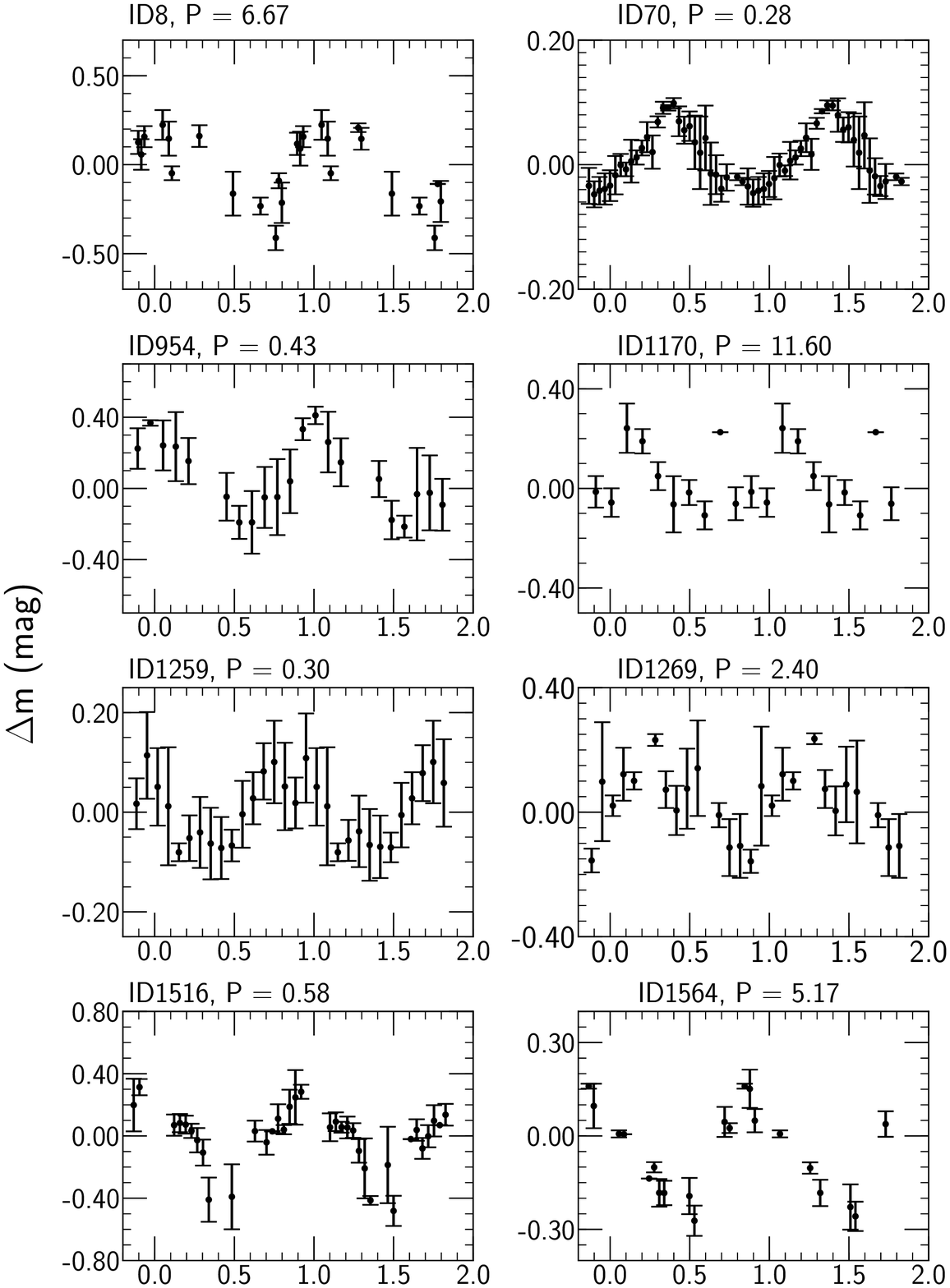}
 \includegraphics[width=8 cm,height=8.0cm]{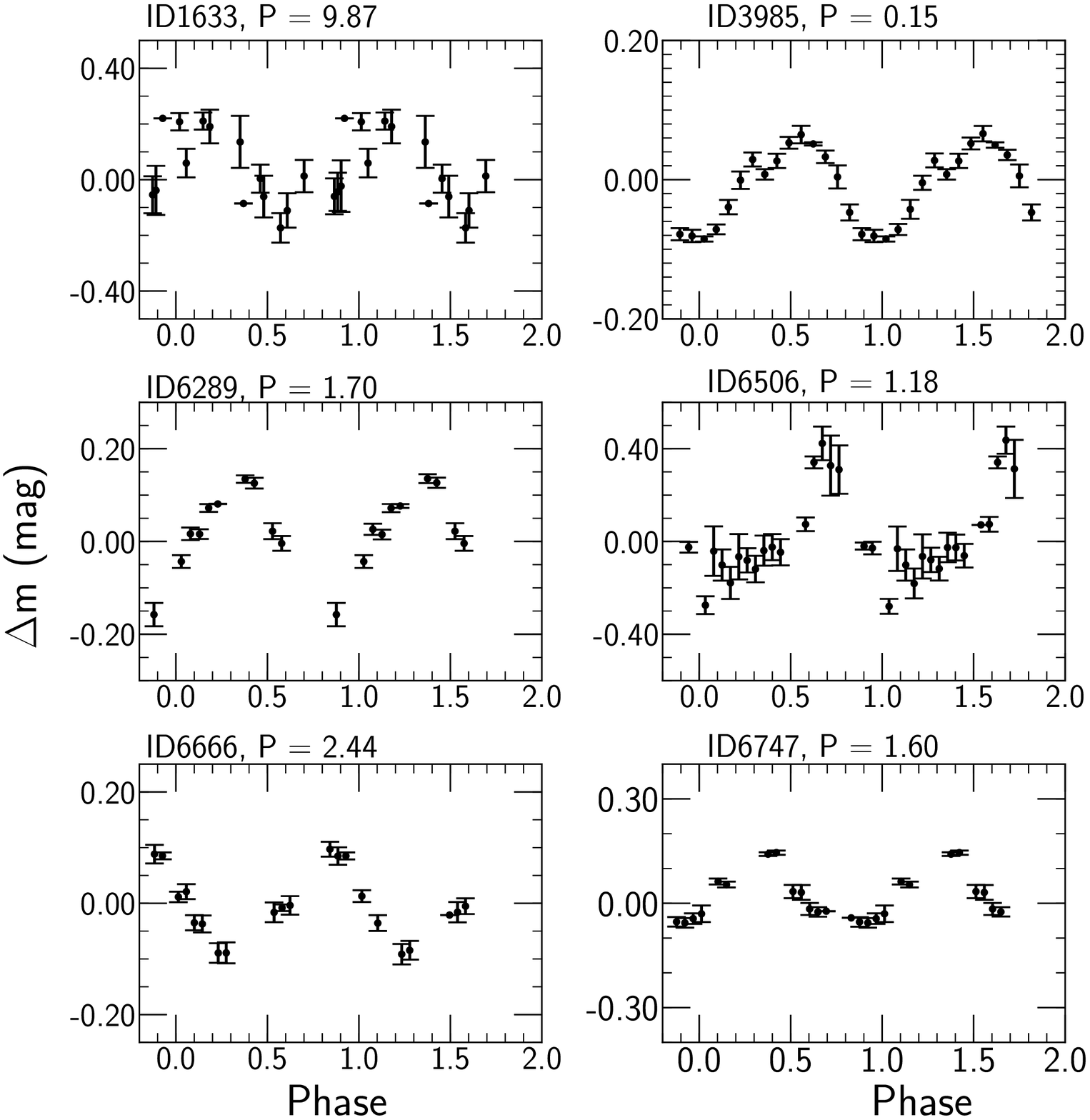}
  \caption{Phase light curves of periodic variables (see text for details).}
  \label{fig:phase}
\end{figure}

We used the LS algorithm at the Starlink\footnote{http://starlink.eao.hawaii.edu/starlink} software package. 
Our time-series data consists of 17 night sparse data, where each night observations are relatively dense packed. We feed all the points as a single file to the software package, further we verified the small periodic variable with each night observations separately in order to avoid any misinterpretation of such unevenly shaped data sets by the LS code. We note that non-variables are often subjected to 0.5 or 1.0 days periodicity in LS analyses, likely due to $\sim$ 1 day separation in our consecutive night observations. Thus, we have rejected certain windows of periods between 0.45 - 0.55 days and 0.90 - 1.1 days to avoid the induced aliasing in our data set. Since the LS method favours sinusoidal light curves, it can not accurately analyse the sharp eclipsing system. However, our variable list does not have many sharp eclipsing systems to draw special attention.
The periods were further verified with period04\footnote{https://www.univie.ac.at/tops/Period04/}\cite[]{2005CoAst.146...53L} and NASA Exoplanet Archive Periodogram Service \footnote{http://exoplanetarchive.ipac.caltech.edu/cgi-bin/Periodogram/nph-simpleupload}. Figure~\ref{fig:freq} displays example LS power spectra of ID70. The estimated periods using the Starlink package are 0.28 days. 
 We estimated the periodicity of 14 stars from our sample, which were all identified as variable from their observed relative photometry light curve. Each observed light curve was folded with the estimated period to generate phase light curve.  Figure \ref{fig:phase} shows  the mean of differential magnitudes in 0.05 phase bin as a function of phase, where the error bars represent the $\sigma$ of the mean in each bin. 
 Out of 31, 17 candidate variables were considered as irregular based on their insufficiently convincing phased light curves.  The periods of all the stars are listed in Table~\ref{tab:parameters}.

\subsection{Previously Known variables} 

 \citet[][]{2012ApJ...755...65R} and \citet[][]{2013AJ....145..113W} detected many variable stars including disk-bearing young stars using the $J$, $H$, $K$ monitoring of the dark clouds Lynds 1003/1004 in the Cyg OB7. Our observing FOV coincides with a part of their coverage. From the list of the variables in \citet[][]{2012ApJ...755...65R} only two (RWA 4 or ID 1771 and RWA 16 or ID 1564) were measured in our field and bright enough to analyze. Since RWA stars were measured in $JHK$ and most are somewhat obscured, hence they were too faint in $I$ band. The variability of those RWA 4 and RWA 16 was confirmed.
  \citet[][]{2012ApJ...755...65R} found ID~1564 and ID~1771 as periodic variables with periodicity 4.84 and 6.35 days, respectively. Our present data suggest a period of 5.17 day for ID1564, whereas ID1771 is aperiodic.  Similarly, Out of 13 variable stars identified by \citet[][]{2013AJ....145..113W} from NIR $JHK$ photometry in this direction of Cygnus OB7, the star ID~70 is studied and found as a variable in the present analyses.  We estimated a period of 0.28 day for ID~70 (RA$_{J2000}$ = $20^h58^m01^s.70$, Dec$_{J2000}$ = $+52^020^m09^s$), which is nicely match with the period (0.30 days; an eclipsing system) measured by \citet[][]{2013AJ....145..113W}. ID~1648 and ID~6666 are listed in the American Association of Variable Star Observers (AAVSO) database from Photometric All Sky Survey (APASS) DR9 \citep[][]{2016yCat.2336....0H}.

\section{Discussions} 
\label{sec:discussion}

\subsection{Pre-main Sequence Variable Stars}

 Photometric variability is a defining characteristic of PMS stars, however, their detection in optical could be limited due to the dominance of circumstellar disk emission in IR. Optical monitoring  has been widely used to characterize the PMS variability \citep[e.g.,][]{2000AJ....120..349H,2001AJ....121.3160C,2004A&A...417..557L,2004AJ....127.2228M,2005MNRAS.358..341L,2005AJ....129..907B,2017ApJ...836...41C,2018MNRAS.476.2813D}. In addition, variability might be a sign of being PMS, but is more often associated with other phenomena, binarity and pulsation related to age. To detect young stars from the variable candidates, we have utilized NIR ($J$, $H$, $K$) and MIR (WISE 3.4, 4.6, 12 and 22 $\mu$m).

\subsubsection{Spectral Energy distribution}
\label{sec:SED_text}
The evolutionary stages of the YSOs could be suggested by the spectral energy distribution (SED). We estimated spectral index ($\alpha = d \log  (\lambda F_\lambda)/d \log(\lambda)$) of the variable sources. Following \cite{2006AJ....131.1574L}, objects with $\alpha$ $\geq$ +0.3 are considered as Class I, $-0.3 \geq \alpha \geq +0.3$ flat spectrum, $-0.3 < \alpha < -1.8$ as Class II and $\alpha < -1.8$ as Class III sources. Out of 31 variable stars, two have WISE measurements in all four bands. We estimated $\alpha$ for these sources from a least-square fit to the fluxes in the range of 3.4 to 24 $\mu$m. The $\alpha$ indices of 10 sources are obtained from available $K$  to 12 $\mu$m. 
  Other 19 sources are not considered for SED analysis in the present discussion since they do not have longer wavelength coverage, whereas  $K$ to $4.6$ $\mu$m have very little leverage and are very susceptible to the reddening in comparison with $K$ to 12 $\mu$m and 3.4 to 24 $\mu$m.
 All the 12  estimated $\alpha$ indices are listed in Table~\ref{tab:parameters}. Using the above approach, we find that four variables  (ID750, ID1564, ID1771, and ID 6506) have $\alpha > -1.8$, which are probable CTTSs. Other eight are either WTTS/Class III, main-sequence (MS) or other field stars. Since, there is no well-defined boundary between WTTSs  and field stars, based on SED analyses we could not classify them. We consider the remaining unexplored 19 variables as field stars or diskless PMS stars. However, high resolution multiwavelength spectroscopic observations could further narrow down such classification \citep[][]{1997AJ....113.1733H,2012ApJ...755...65R,2004ApJ...602..816L,2014ApJ...786...97H,2015MNRAS.454.3597D}.

\subsubsection{NIR Colour-Colour and Colour-Magnitude Diagram}

\begin{figure}
\includegraphics[width=8.0 cm,height=8.0 cm]{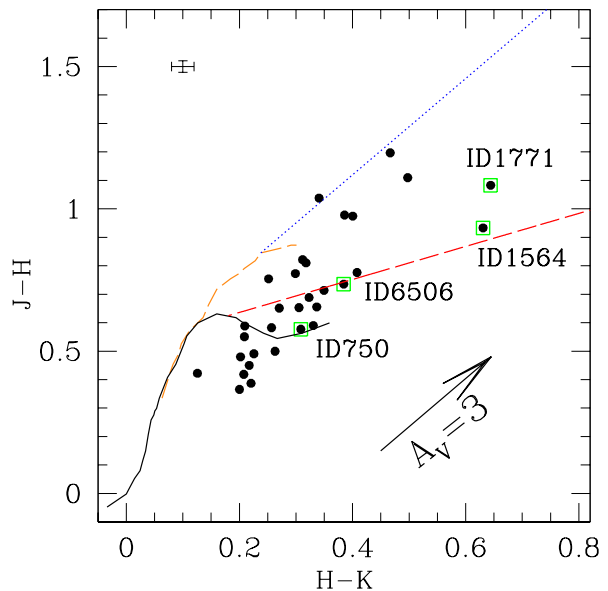}
\caption{The $J-H/H-K$ CC diagram of candidate variable stars in Cyg OB7. The green boxes represent the location of the probable CTTSs identified using $\alpha$ indices.  The locus for dwarfs (solid black) and giants (long dashed golden line) are taken from Bessel $\&$ Brett (1988). The long dashed red line represents the CTTSs locus from Meyer et al. (1997). Average error bars are shown in the left top corner. The dotted blue line is the reddening vector taken from  Cohen et al. 1981. The reddening vector of visual extinction $A_V$ = 3 mag is also marked (see text).}
  \label{fig:jhk_var} 
\end{figure}

The NIR colour-colour (CC) diagram is shown in Figure~\ref{fig:jhk_var} for all the identified candidate variable stars. The locus of dwarfs (solid black line) and giants (long dashed golden line) are adopted from \citet{1988PASP..100.1134B}. The long dashed red line represents the linear fitting of observed CTTSs by \citet{1997AJ....114..288M}. The dotted blue line indicates a reddening vector with slope $A_J$/$A_V$ = 0.265, $A_H$/$A_V$ = 0.155 and $A_K$/$A_V$ = 0.090 taken from \cite{1981ApJ...249..481C}, a representative reddening vector of $A_V$ = 3 mag is also shown in arrow mark. The NIR emission of normal stars is thought to originate from the photosphere and they follow the dwarf locus in NIR CC diagram, whereas their giant counterparts appear more luminous. Nevertheless, the NIR emission of CTTSs is dominated by the circumstellar disk in addition to photospheric emission \citep{1992ApJ...393..278L}. Thus, they advance according to the reddening vector based on the amount of emission coming from the circumstellar material.

 Objects ID1564 and 1771 are both absorbed by over A$_V$ = 3 mag of extinction in Figure~\ref{fig:jhk_var}. Additionally, from our SED analyses, we also find that ID740 and ID 6506 are CTTSs. All CTTSs are marked with the green boxes in Figure~\ref{fig:jhk_var}. The remaining 27 are either field stars or diskless PMS/Class III stars. Naturally, Class III sources are very difficult to distinguish from the field stars based on NIR excess. We suggest further high-resolution spectroscopic observations to evaluate physically each of the sources.

The stars ID1564 and ID1771 also show strong H$_\alpha$ emission (equivalent width $>$ 10 ${\AA}$) in ($r-i$)/($r-H_\alpha$) CC diagram (Figures are not shown) based on IPHAS photometry  \citep[e.g.,][]{2011MNRAS.415..103B, 2018MNRAS.476.2813D}.

\begin{figure}
\includegraphics[width=8.0 cm,height=8.0 cm]{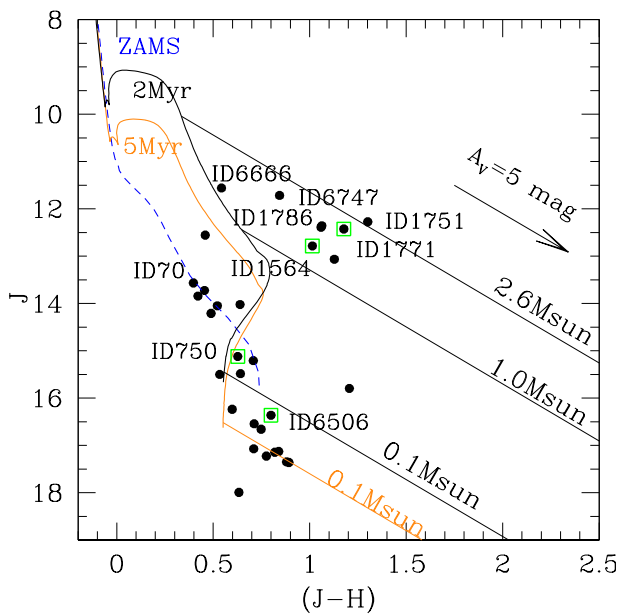}
  \caption{The $J$/($J-H$) CMD for all the candidate variable stars. The green boxes are the CTTSs, the identification is marked. All the stars with age $<$ 5 Myr (see Figure \ref{fig:viv_yso})  are also marked. The blue curve indicates the ZAMS. Two representative age 2 and 5 Myr are shown in black and yellow, respectively. Different reddening vectors are drawn from 2.6, 1.0, 0.1 M$\sun$ mass for different ages. All the isochrones and tracks are corrected for the distance. A representative reddening vector of visual extinction $A_V$ = 5 mag is also displayed.}
  \label{fig:jhj_yso}
\end{figure}

The $J$ vs. ($J-H$) CMD is another important tool to get the approximate knowledge about masses of the PMSs, since they are more easily detectable over more extinct optical (e.g. $V$ vs. $V-I$) wavelengths from the molecular cloud. Furthermore, the longer wavelengths (e.g. $K$ vs $H-K$) CMD might be contaminated by circumstellar extinction. In Figure \ref{fig:jhj_yso}, the candidate variables are compared with mass bins of representative ages 2 and 5 Myr. Notably, the location of PMSs moves along the reddening vector from the theoretical isochrones depending upon circumstellar disk emission. Two PMS variables (ID1564 and ID1771) seem to confine between 2.6-1.0 M$\sun$, another two (ID750 and ID6506) appear to be low mass stars ($\sim$ 0.1 M$\sun$).  Since we have used PMS isochrone in these analyses, we could not predict the mass range of non-PMS stars. 

This analysis is highly dependent on the age of PMSs, a change between 2$-$5 Myr would result in 30$-$50\% agreement of the present mass estimation. Such mass estimation could also be affected by several errors like the  variable nature of stars, the presence of eclipsing binary, emission from the circumstellar material.

\subsubsection{Optical Colour-Magnitude Diagram}
\label{sec:viv_cmd}
\begin{figure}
\includegraphics[width=8.0 cm,height=8.0 cm]{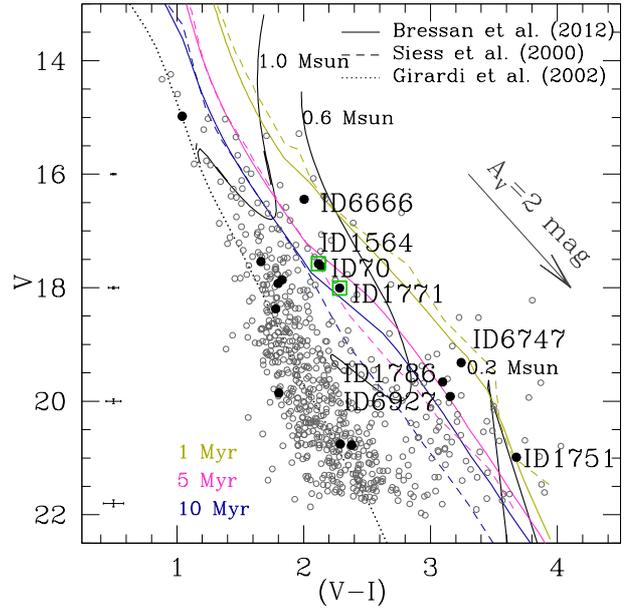}
  \caption{The $V/(V-I)$ CMD for all the sources of the studied region. The black circles are the candidate variable sources, whereas green boxes are CTTSs (the identification are marked). The identification of the stars with age $<$ 5 Myr is also marked. The dotted curve is the locus of ZAMS from Girardi et al. (2002), solid curves are the PMS isochrones of age 1.0, 5.0 and 10.0 Myr, respectively, and the thin black solid curves are the evolutionary tracks for various mass bins from Siess et al. (2000). The long dashed curves are PMS isochrones of age 1.0, 5.0 and 10.0 Myr taken from Siess et al. (2000). All the isochrones and tracks are corrected for the distance and reddening. Average error bars are shown on the left side. The reddening vector of visual extinction $A_V$ = 2 mag is shown.}
  \label{fig:viv_yso}
\end{figure}

The observed $VI$ photometry was employed to estimate the approximate ages and masses of PMS candidate variables. The $V$ versus $(V-I)$ Colour-Magnitude Diagram (CMD) is displayed in Figure~\ref{fig:viv_yso}, where {\it zero-age main sequence} (ZAMS) from \citet[][]{2002A&A...391..195G} is represented by dotted black curve.  Different theoretical PMS isochrones \citep[][]{2012MNRAS.427..127B,2000A&A...358..593S} for ages of 1, 5 and 10 Myr are shown as the nearly vertical lines. The solid black curves are evolutionary mass tracks for different mass bins from \citet[][]{2000A&A...358..593S}. All the theoretical models are corrected for cluster distance 760~pc\footnote{The distance of the region is found to be in the range 600 - 800 pc using Gaia data release 2, which is also in agreement with  this assumed distance.}. The background stars are fitted well with a reddening of $E(B-V)$ = 0.30 mag, which is considered as background extinction of the studied region. This background  reddening value in optical is also consistent with the single epoch UKIDSS (NIR) data.

We estimated the masses and ages of a few PMS candidate variable stars using interpolation methods of \citet[][]{2000A&A...358..593S} isochrones in the CMD. The estimated ages and masses are listed in Table \ref{tab:parameters}.  Out of 31, 15 variables have V-band detection (black circles in Figure \ref{fig:viv_yso}). 
The CMD positions of ID1564 and ID1771 CTTS variables seem to be adequately fit for $\sim$ 5 Myr, another six sources have age less than 5 Myr. Although, different models at the low-mass end differ significantly as we can see in Fig.~\ref{fig:viv_yso}. The masses of the variables seem to be in the range 0.2$-$1.0 M$_\odot$. Notably, eight sources are located along ZAMS. Since we are using PMSs isochrone in these analyses, the stars located around ZAMS are not considered for mass and age estimation in the above approach. The reddening vector is nearly parallel to the isochrones, a small extinction variation would not have much effect on the age estimation of variable stars. However, such measurement will lead to 50\%-60\% error in case of unresolved binary. The presence of binary will brighten a star from its actual measurement, hence a lower age estimation is predicted. Nonetheless, the average age of young stars is deduced 1$-$2 Myr \citep[e.g.,][]{2009ApJS..184...18G}. The extinction from the natal molecular environment and circumstellar envelope around young stellar objects could introduce over local background subtraction during photometry, which could lead to an apparent age spread up to $\sim$ 10 Myr in optical CMD.

\begin{figure}
\includegraphics[width=8.0 cm,height=8.0 cm]{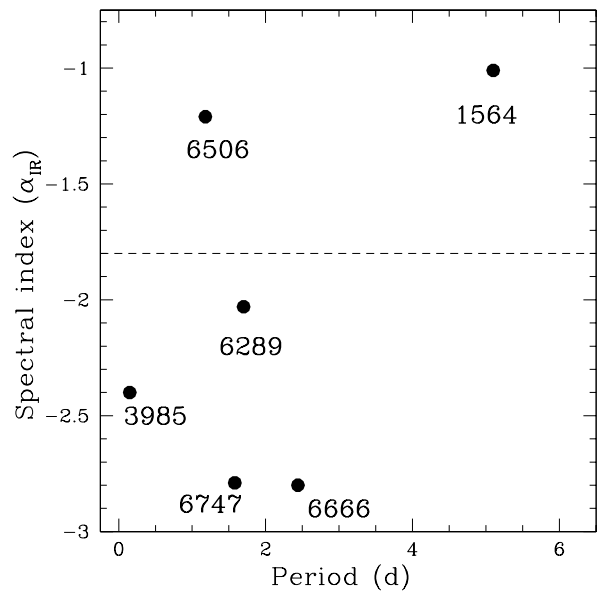}
  \caption{Spectral indices of the candidate variable stars, as defined by Lada et al. (2006), on the basis of their IR data, shown as a function of the rotational period. The dashed line at $\alpha$= -1.8 represent the separation between probable CTTSs and diskless stars. All the star identifications are marked (see text for details).
 }
  \label{fig:alpha_period}
\end{figure}

\subsection{Other variables: Eclipsing binaries}
 As discussed earlier 13 \% (4 out of 31) variables are PMS. The rest are field stars, with some diskless PMS stars. We have estimated the periodicity of 45 \% (14 out of 31) variable candidates. Remaining 55 \% are considered stochastic sources, however, only two (ID1771 and ID750) among them have infrared excess. The variable stars could be classified based on the shape of their phase and observed light curves. Note that we do not attempt to classify the light curves by the physical characteristics of the stars due to the lack of sufficient data.

Eclipsing binaries are easiest detectable variable stars due to their smooth, stable periodic light curves, prominent sudden drop on their magnitudes. The smooth phase light curves of five sources (ID70, ID3985, ID6289, ID6666, ID6747) indicate that they are an eclipsing system. The star ID70 was earlier detected as W UMa (or WMa) eclipsing binary system with a short period (0.30 days) by \citet[][]{2013AJ....145..113W}. Short periods and smooth observed light curve pattern of ID3985 is also similar to  WMa objects. This type of eclipsing binary variable is known as a low mass contact binary of stars with spectral type G or K. The stars ID8, ID1170, ID6506 are suspected to be $\delta$-scuti. ID1259 is probably a young eruptive variable of type Ex-Lupi, however, we do not find any signature of infrared excess around it in this analyses (see also section \ref{sec:SED_text}).

\subsubsection{Correlation between photometric variability and circumstellar disks}

 Since circumstellar disks play a vital role in PMS variability \citep[e.g.,][]{1993AJ....106..372E,2002A&A...396..513H,2006AJ....132.1555N,2018MNRAS.476.2813D}, we looked for any possible correlation between NIR excess and rotation.
Figure \ref{fig:alpha_period} displays spectral index against rotation period. While the number of stars is too small to provide a good statistical significance, the general form of the distribution is quite similar to what is seen in the ONC \cite[][]{2002A&A...396..513H}, IC 348 \cite[][]{2006AJ....132.1555N} or NGC 2282 \citep[e.g.,][]{2018MNRAS.476.2813D}. 
 One CTTS star (ID6506) has the period less than 2 days and another CTTS (ID1564) is relatively slow rotators.
 A small number of rotators in our sample prevents to establish any conclusion, notably, the rapid rotators tend to have less infrared excess (larger negative slopes), while the slower rotators should have a mixture of stars with optically thick accretion disks and stars without such disks. More interestingly, all the eclipsing binaries have a small period and large spectral indices. Such large values of alpha indices of binaries, beyond the blackbody ($\alpha_{BB}$ = -1.80), possibly arising from the contamination of photometry by the binary counterpart, which make them more luminous as they are. However, such steeper $\alpha_{BB}$ could also be driven by bad data at the red end.

\begin{figure*}
\includegraphics[width=16.0 cm,height=9.0 cm]{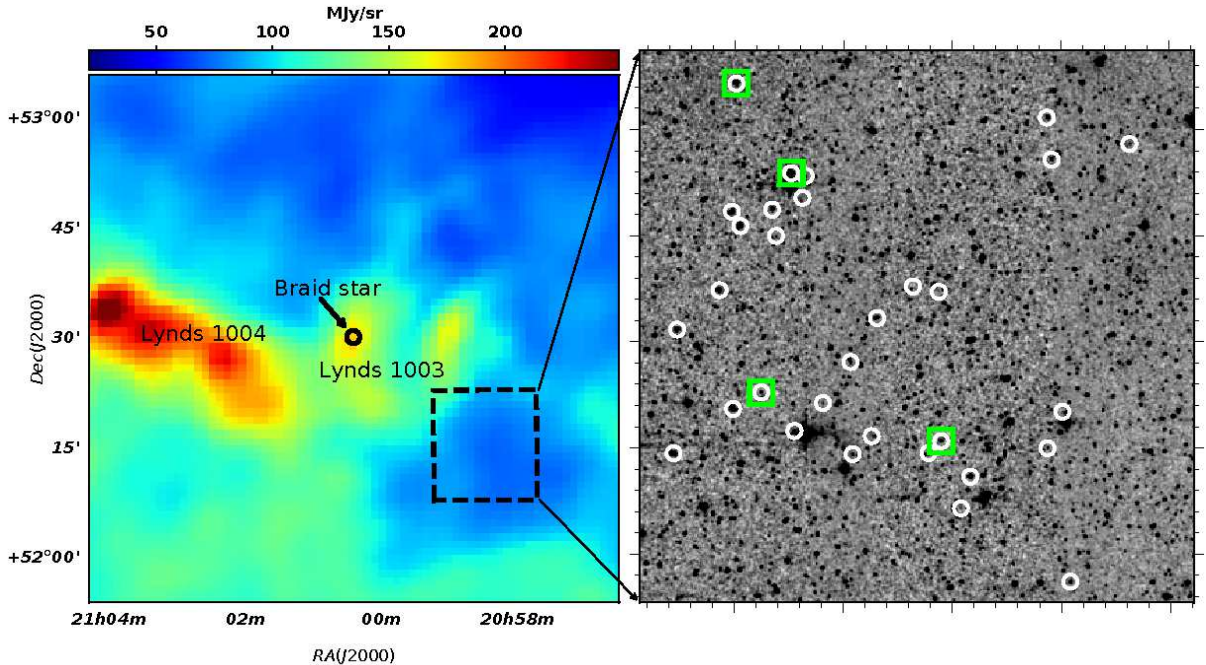}
  \caption{({\it left panel}) Planck 857 GHz map towards the Lynds 1003/1004 molecular clouds. Our studied region is marked with black box. ({\it right panel}) Spatial distribution of all the  variable sources is over plotted on 2MASS $K_s$ image. Green sources are CTTSs (see text for details).}
  \label{fig:planck_iras_spatial}
\end{figure*}

\subsubsection{Large-scale view of spatial distribution of variables} 
 
Figure \ref{fig:observed_field_spatial} ({\it left panel}) shows the large-scale structure around dark cloud Lynds 1003/1004 in IR (2mass K: blue; WISE W1: green; WISE W3: red). The WISE W3 (12 $\mu$m) band is affected by the PAH emissions, which are believed to be excited by UV radiation from the massive star(s) and are present at the interface between the \hii and neutral molecular gas.  In Figure \ref{fig:planck_iras_spatial} ({\it left panel}), we displayed the studied region in Planck 857 GHz (350 $\mu$m) map. As the high-frequency channels of the third generation mission, Planck covers the peak thermal emission frequencies of dust colder than 14 K, therefore, Planck images probe the coldest parts of the ISM \citep[][]{2011A&A...536A...1P}. Thus, Planck images are very useful to correlate between the PMSs in the star-forming regions \citep[e.g.,][]{2018ApJ...864..154D}.  
 
 In Figure \ref{fig:planck_iras_spatial} ({\it right panel}), the spatial distribution of candidate variables is shown on  2MASS $K_s$ images. Two PMS stars (ID1564, 1771)are located in the northeastern direction, these are probably members of Lynds 1003 cloud. Another two PMSs (ID750, 6506) is also along the faint dust ridge in the south eastern direction.

\section{Summary and Conclusions}
\label{sec:summary}
In this paper, we presented $I$ band time-series photometry of stars towards Cyg OB7 complimented with the  optical ($V,R, I$) observations, archival NIR ($J, H, K$), WISE (3.4, 4.6, 12 and 22 $\mu$m), and Planck (857 GHz) data.  We summarized the main results as follows:

 \begin{enumerate}

  \item The $I$-band (down to $\sim$ 20.5 mag) light curve analysis enables us to probe the low mass variable stars, including PMSs.  From the photometric light curves, we identified 31 photometric variable stars towards the region. Out of 31, 5 are known from earlier studies, 26 are new variables from this study. The periods of 14 variables are estimated to be in the range of 0.15 to 11.60 days.

  \item From SED analysis and location on $J-H$/$H-K$ CC diagram, we selected four variable stars are CTTSs. The remaining stars are unclassified in this study.

  \item The $V$/($V-I$) CMD suggested that variable CTTSs have age $<$ 5 Myr and the masses in the range 0.2$-$1.0 M$_\odot$.

  \item Five eclipsing systems are identified from the light curves, including one previously known eclipsing binary.

 \end{enumerate}

In the beginning, the goal of this project was to relate infrared excess with variability. Unfortunately, a small number of CTTS variables does not represent any obvious correlation. However, we have identified four CTTSs including two previously identified. Photometric monitoring of more number of SFRs would be able to represent a statistically significant correlation between photometric variability and circumstellar disk.

\section*{Acknowledgments}
We are thankful to Dr. Scott Wolk for the valuable comments, which helps us to improve the overall quality of the paper. This work is supported by Satyendra Nath Bose National Centre for Basic Sciences, Kolkata, India under the Department of Science and Technology (DST), Govt. of India.  We would like to thank Prof. Wen-Ping Chen for valuable suggestions.  This paper made use of the data obtained as part of the UKIRT Infrared Deep Sky Survey (UKIDSS). This publication also used the data products from the SIMBAD database (operated at CDS, Strasbourg, France), the Two Micron All Sky Survey, which is a joint project of the University of Massachusetts and the Infrared Processing and Analysis Center/California Institute of Technology, funded by NASA and NSF, archival data obtained with the Wide-Field Infrared Survey Explorer (a joint project of the University of California, Los Angeles, and the Jet Propulsion Laboratory [JPL], California Institute of Technology [Caltech], funded by the National Aeronautics and Space Administration [NASA]), and the NOAO Science archive, which is operated by the Association of Universities for  Research in Astronomy (AURA), Inc., under a cooperative agreement with the National Science Foundation. 

\bibliographystyle{mnras}
\bibliography{cygob7}

\begin{thebibliography}{}
\makeatletter
\relax
\def\mn@urlcharsother{\let\do\@makeother \do\$\do\&\do\#\do\^\do\_\do\%\do\~}
\def\mn@doi{\begingroup\mn@urlcharsother \@ifnextchar [ {\mn@doi@}
  {\mn@doi@[]}}
\def\mn@doi@[#1]#2{\def\@tempa{#1}\ifx\@tempa\@empty \href
  {http://dx.doi.org/#2} {doi:#2}\else \href {http://dx.doi.org/#2} {#1}\fi
  \endgroup}
\def\mn@eprint#1#2{\mn@eprint@#1:#2::\@nil}
\def\mn@eprint@arXiv#1{\href {http://arxiv.org/abs/#1} {{\tt arXiv:#1}}}
\def\mn@eprint@dblp#1{\href {http://dblp.uni-trier.de/rec/bibtex/#1.xml}
  {dblp:#1}}
\def\mn@eprint@#1:#2:#3:#4\@nil{\def\@tempa {#1}\def\@tempb {#2}\def\@tempc
  {#3}\ifx \@tempc \@empty \let \@tempc \@tempb \let \@tempb \@tempa \fi \ifx
  \@tempb \@empty \def\@tempb {arXiv}\fi \@ifundefined
  {mn@eprint@\@tempb}{\@tempb:\@tempc}{\expandafter \expandafter \csname
  mn@eprint@\@tempb\endcsname \expandafter{\@tempc}}}

\bibitem[\protect\citeauthoryear{{Aspin} et~al.,}{{Aspin}
  et~al.}{2008}]{2008arXiv0810.3943A}
{Aspin} C.,  et~al., 2008, preprint, \href
  {http://adsabs.harvard.edu/abs/2008arXiv0810.3943A} {} (\mn@eprint {arXiv}
  {0810.3943})

\bibitem[\protect\citeauthoryear{{Barentsen} et~al.,}{{Barentsen}
  et~al.}{2011}]{2011MNRAS.415..103B}
{Barentsen} G.,  et~al., 2011, \mn@doi [\mnras]
  {10.1111/j.1365-2966.2011.18674.x}, \href
  {http://adsabs.harvard.edu/abs/2011MNRAS.415..103B} {415, 103}

\bibitem[\protect\citeauthoryear{{Bessell} \& {Brett}}{{Bessell} \&
  {Brett}}{1988}]{1988PASP..100.1134B}
{Bessell} M.~S.,  {Brett} J.~M.,  1988, \mn@doi [\pasp] {10.1086/132281}, \href
  {http://adsabs.harvard.edu/abs/1988PASP..100.1134B} {100, 1134}

\bibitem[\protect\citeauthoryear{{Bouvier}, {Cabrit}, {Fernandez}, {Martin}  \&
  {Matthews}}{{Bouvier} et~al.}{1993}]{1993A&A...272..176B}
{Bouvier} J.,  {Cabrit} S.,  {Fernandez} M.,  {Martin} E.~L.,   {Matthews}
  J.~M.,  1993, \aap, \href
  {http://adsabs.harvard.edu/abs/1993A%26A...272..176B} {272, 176}

\bibitem[\protect\citeauthoryear{{Bressan}, {Marigo}, {Girardi}, {Salasnich},
  {Dal Cero}, {Rubele}  \& {Nanni}}{{Bressan}
  et~al.}{2012}]{2012MNRAS.427..127B}
{Bressan} A.,  {Marigo} P.,  {Girardi} L.,  {Salasnich} B.,  {Dal Cero} C.,
  {Rubele} S.,   {Nanni} A.,  2012, \mn@doi [\mnras]
  {10.1111/j.1365-2966.2012.21948.x}, \href
  {http://adsabs.harvard.edu/abs/2012MNRAS.427..127B} {427, 127}

\bibitem[\protect\citeauthoryear{{Brice{\~n}o}, {Calvet}, {Hern{\'a}ndez},
  {Vivas}, {Hartmann}, {Downes}  \& {Berlind}}{{Brice{\~n}o}
  et~al.}{2005}]{2005AJ....129..907B}
{Brice{\~n}o} C.,  {Calvet} N.,  {Hern{\'a}ndez} J.,  {Vivas} A.~K.,
  {Hartmann} L.,  {Downes} J.~J.,   {Berlind} P.,  2005, \mn@doi [\aj]
  {10.1086/426911}, \href {http://adsabs.harvard.edu/abs/2005AJ....129..907B}
  {129, 907}

\bibitem[\protect\citeauthoryear{{Carpenter}, {Hillenbrand}  \&
  {Skrutskie}}{{Carpenter} et~al.}{2001}]{2001AJ....121.3160C}
{Carpenter} J.~M.,  {Hillenbrand} L.~A.,   {Skrutskie} M.~F.,  2001, \mn@doi
  [\aj] {10.1086/321086}, \href
  {http://adsabs.harvard.edu/abs/2001AJ....121.3160C} {121, 3160}

\bibitem[\protect\citeauthoryear{{Cody}, {Hillenbrand}, {David}, {Carpenter},
  {Everett}  \& {Howell}}{{Cody} et~al.}{2017}]{2017ApJ...836...41C}
{Cody} A.~M.,  {Hillenbrand} L.~A.,  {David} T.~J.,  {Carpenter} J.~M.,
  {Everett} M.~E.,   {Howell} S.~B.,  2017, \mn@doi [\apj]
  {10.3847/1538-4357/836/1/41}, \href
  {http://adsabs.harvard.edu/abs/2017ApJ...836...41C} {836, 41}

\bibitem[\protect\citeauthoryear{{Cohen}}{{Cohen}}{1980}]{1980AJ.....85...29C}
{Cohen} M.,  1980, \mn@doi [\aj] {10.1086/112630}, \href
  {http://adsabs.harvard.edu/abs/1980AJ.....85...29C} {85, 29}

\bibitem[\protect\citeauthoryear{{Cohen}, {Persson}, {Elias}  \&
  {Frogel}}{{Cohen} et~al.}{1981}]{1981ApJ...249..481C}
{Cohen} J.~G.,  {Persson} S.~E.,  {Elias} J.~H.,   {Frogel} J.~A.,  1981,
  \mn@doi [\apj] {10.1086/159308}, \href
  {http://adsabs.harvard.edu/abs/1981ApJ...249..481C} {249, 481}

\bibitem[\protect\citeauthoryear{{Cutri} \& {et al.}}{{Cutri} \& {et
  al.}}{2012}]{2012yCat.2311....0C}
{Cutri} R.~M.,  {et al.} 2012, VizieR Online Data Catalog, \href
  {http://adsabs.harvard.edu/abs/2012yCat.2311....0C} {2311}

\bibitem[\protect\citeauthoryear{{Cutri} et~al.,}{{Cutri}
  et~al.}{2003}]{2003yCat.2246....0C}
{Cutri} R.~M.,  et~al., 2003, VizieR Online Data Catalog, \href
  {http://adsabs.harvard.edu/abs/2003yCat.2246....0C} {2246}

\bibitem[\protect\citeauthoryear{{Devine}, {Reipurth}  \& {Bally}}{{Devine}
  et~al.}{1997}]{1997IAUS..182P..91D}
{Devine} D.,  {Reipurth} B.,   {Bally} J.,  1997, in {Reipurth} B.,  {Bertout}
  C.,  eds,  IAU Symposium Vol. 182, Herbig-Haro Flows and the Birth of Stars.
  p.~91

\bibitem[\protect\citeauthoryear{{Dutta}, {Mondal}, {Jose}, {Das}, {Samal}  \&
  {Ghosh}}{{Dutta} et~al.}{2015}]{2015MNRAS.454.3597D}
{Dutta} S.,  {Mondal} S.,  {Jose} J.,  {Das} R.~K.,  {Samal} M.~R.,   {Ghosh}
  S.,  2015, \mn@doi [\mnras] {10.1093/mnras/stv2190}, \href
  {http://adsabs.harvard.edu/abs/2015MNRAS.454.3597D} {454, 3597}

\bibitem[\protect\citeauthoryear{{Dutta}, {Mondal}, {Joshi}, {Jose}, {Das}  \&
  {Ghosh}}{{Dutta} et~al.}{2018a}]{2018MNRAS.476.2813D}
{Dutta} S.,  {Mondal} S.,  {Joshi} S.,  {Jose} J.,  {Das} R.,   {Ghosh} S.,
  2018a, \mn@doi [\mnras] {10.1093/mnras/sty329}, \href
  {http://adsabs.harvard.edu/abs/2018MNRAS.476.2813D} {476, 2813}

\bibitem[\protect\citeauthoryear{{Dutta}, {Mondal}, {Samal}  \& {Jose}}{{Dutta}
  et~al.}{2018b}]{2018ApJ...864..154D}
{Dutta} S.,  {Mondal} S.,  {Samal} M.~R.,   {Jose} J.,  2018b, \mn@doi [\apj]
  {10.3847/1538-4357/aadb3e}, \href
  {http://adsabs.harvard.edu/abs/2018ApJ...864..154D} {864, 154}

\bibitem[\protect\citeauthoryear{{Edwards} et~al.,}{{Edwards}
  et~al.}{1993}]{1993AJ....106..372E}
{Edwards} S.,  et~al., 1993, \mn@doi [\aj] {10.1086/116646}, \href
  {http://adsabs.harvard.edu/abs/1993AJ....106..372E} {106, 372}

\bibitem[\protect\citeauthoryear{{Girardi}, {Bertelli}, {Bressan}, {Chiosi},
  {Groenewegen}, {Marigo}, {Salasnich}  \& {Weiss}}{{Girardi}
  et~al.}{2002}]{2002A&A...391..195G}
{Girardi} L.,  {Bertelli} G.,  {Bressan} A.,  {Chiosi} C.,  {Groenewegen}
  M.~A.~T.,  {Marigo} P.,  {Salasnich} B.,   {Weiss} A.,  2002, \mn@doi [\aap]
  {10.1051/0004-6361:20020612}, \href
  {http://adsabs.harvard.edu/abs/2002A%26A...391..195G} {391, 195}

\bibitem[\protect\citeauthoryear{{Gutermuth}, {Megeath}, {Myers}, {Allen},
  {Pipher}  \& {Fazio}}{{Gutermuth} et~al.}{2009}]{2009ApJS..184...18G}
{Gutermuth} R.~A.,  {Megeath} S.~T.,  {Myers} P.~C.,  {Allen} L.~E.,  {Pipher}
  J.~L.,   {Fazio} G.~G.,  2009, \mn@doi [\apjs] {10.1088/0067-0049/184/1/18},
  \href {http://adsabs.harvard.edu/abs/2009ApJS..184...18G} {184, 18}

\bibitem[\protect\citeauthoryear{{Henden}, {Templeton}, {Terrell}, {Smith},
  {Levine}  \& {Welch}}{{Henden} et~al.}{2016}]{2016yCat.2336....0H}
{Henden} A.~A.,  {Templeton} M.,  {Terrell} D.,  {Smith} T.~C.,  {Levine} S.,
  {Welch} D.,  2016, VizieR Online Data Catalog, \href
  {http://adsabs.harvard.edu/abs/2016yCat.2336....0H} {2336}

\bibitem[\protect\citeauthoryear{{Herbig}}{{Herbig}}{1962}]{1962AdA&A...1...47H}
{Herbig} G.~H.,  1962, Advances in Astronomy and Astrophysics, \href
  {http://adsabs.harvard.edu/abs/1962AdA%26A...1...47H} {1, 47}

\bibitem[\protect\citeauthoryear{{Herbig} \& {Bell}}{{Herbig} \&
  {Bell}}{1988}]{1988cels.book.....H}
{Herbig} G.~H.,  {Bell} K.~R.,  1988, {Third Catalog of Emission-Line Stars of
  the Orion Population : 3 : 1988}

\bibitem[\protect\citeauthoryear{{Herbst}, {Maley}  \& {Williams}}{{Herbst}
  et~al.}{2000}]{2000AJ....120..349H}
{Herbst} W.,  {Maley} J.~A.,   {Williams} E.~C.,  2000, \mn@doi [\aj]
  {10.1086/301430}, \href {http://adsabs.harvard.edu/abs/2000AJ....120..349H}
  {120, 349}

\bibitem[\protect\citeauthoryear{{Herbst}, {Bailer-Jones}, {Mundt},
  {Meisenheimer}  \& {Wackermann}}{{Herbst} et~al.}{2002}]{2002A&A...396..513H}
{Herbst} W.,  {Bailer-Jones} C.~A.~L.,  {Mundt} R.,  {Meisenheimer} K.,
  {Wackermann} R.,  2002, \mn@doi [\aap] {10.1051/0004-6361:20021362}, \href
  {http://adsabs.harvard.edu/abs/2002A%26A...396..513H} {396, 513}

\bibitem[\protect\citeauthoryear{{Herczeg} \& {Hillenbrand}}{{Herczeg} \&
  {Hillenbrand}}{2014}]{2014ApJ...786...97H}
{Herczeg} G.~J.,  {Hillenbrand} L.~A.,  2014, \mn@doi [\apj]
  {10.1088/0004-637X/786/2/97}, \href
  {http://adsabs.harvard.edu/abs/2014ApJ...786...97H} {786, 97}

\bibitem[\protect\citeauthoryear{{Hillenbrand}}{{Hillenbrand}}{1997}]{1997AJ....113.1733H}
{Hillenbrand} L.~A.,  1997, \mn@doi [\aj] {10.1086/118389}, \href
  {http://adsabs.harvard.edu/abs/1997AJ....113.1733H} {113, 1733}

\bibitem[\protect\citeauthoryear{{Hiltner}}{{Hiltner}}{1956}]{1956ApJS....2..389H}
{Hiltner} W.~A.,  1956, \mn@doi [\apjs] {10.1086/190029}, \href
  {http://adsabs.harvard.edu/abs/1956ApJS....2..389H} {2, 389}

\bibitem[\protect\citeauthoryear{{Horne} \& {Baliunas}}{{Horne} \&
  {Baliunas}}{1986}]{1986ApJ...302..757H}
{Horne} J.~H.,  {Baliunas} S.~L.,  1986, \mn@doi [\apj] {10.1086/164037}, \href
  {http://adsabs.harvard.edu/abs/1986ApJ...302..757H} {302, 757}

\bibitem[\protect\citeauthoryear{{Joy}}{{Joy}}{1945}]{1945ApJ...102..168J}
{Joy} A.~H.,  1945, \mn@doi [\apj] {10.1086/144749}, \href
  {http://adsabs.harvard.edu/abs/1945ApJ...102..168J} {102, 168}

\bibitem[\protect\citeauthoryear{{Lada} \& {Adams}}{{Lada} \&
  {Adams}}{1992}]{1992ApJ...393..278L}
{Lada} C.~J.,  {Adams} F.~C.,  1992, \mn@doi [\apj] {10.1086/171505}, \href
  {http://adsabs.harvard.edu/abs/1992ApJ...393..278L} {393, 278}

\bibitem[\protect\citeauthoryear{{Lada} et~al.,}{{Lada}
  et~al.}{2006}]{2006AJ....131.1574L}
{Lada} C.~J.,  et~al., 2006, \mn@doi [\aj] {10.1086/499808}, \href
  {http://adsabs.harvard.edu/abs/2006AJ....131.1574L} {131, 1574}

\bibitem[\protect\citeauthoryear{{Lamarre} et~al.,}{{Lamarre}
  et~al.}{2010}]{2010A&A...520A...9L}
{Lamarre} J.-M.,  et~al., 2010, \mn@doi [\aap] {10.1051/0004-6361/200912975},
  \href {http://adsabs.harvard.edu/abs/2010A%26A...520A...9L} {520, A9}

\bibitem[\protect\citeauthoryear{{Lamm}, {Bailer-Jones}, {Mundt}, {Herbst}  \&
  {Scholz}}{{Lamm} et~al.}{2004}]{2004A&A...417..557L}
{Lamm} M.~H.,  {Bailer-Jones} C.~A.~L.,  {Mundt} R.,  {Herbst} W.,   {Scholz}
  A.,  2004, \mn@doi [\aap] {10.1051/0004-6361:20035588}, \href
  {http://adsabs.harvard.edu/abs/2004A%26A...417..557L} {417, 557}

\bibitem[\protect\citeauthoryear{{Landolt}}{{Landolt}}{1992}]{1992AJ....104..340L}
{Landolt} A.~U.,  1992, \mn@doi [\aj] {10.1086/116242}, \href
  {http://adsabs.harvard.edu/abs/1992AJ....104..340L} {104, 340}

\bibitem[\protect\citeauthoryear{{Lata}, {Pandey}, {Chen}, {Maheswar}  \&
  {Chauhan}}{{Lata} et~al.}{2012}]{2012MNRAS.427.1449L}
{Lata} S.,  {Pandey} A.~K.,  {Chen} W.~P.,  {Maheswar} G.,   {Chauhan} N.,
  2012, \mn@doi [\mnras] {10.1111/j.1365-2966.2012.22070.x}, \href
  {http://adsabs.harvard.edu/abs/2012MNRAS.427.1449L} {427, 1449}

\bibitem[\protect\citeauthoryear{{Lenz} \& {Breger}}{{Lenz} \&
  {Breger}}{2005}]{2005CoAst.146...53L}
{Lenz} P.,  {Breger} M.,  2005, \mn@doi [Communications in Asteroseismology]
  {10.1553/cia146s53}, \href
  {http://adsabs.harvard.edu/abs/2005CoAst.146...53L} {146, 53}

\bibitem[\protect\citeauthoryear{{Littlefair}, {Naylor}, {Burningham}  \&
  {Jeffries}}{{Littlefair} et~al.}{2005}]{2005MNRAS.358..341L}
{Littlefair} S.~P.,  {Naylor} T.,  {Burningham} B.,   {Jeffries} R.~D.,  2005,
  \mn@doi [\mnras] {10.1111/j.1365-2966.2005.08737.x}, \href
  {http://adsabs.harvard.edu/abs/2005MNRAS.358..341L} {358, 341}

\bibitem[\protect\citeauthoryear{{Lomb}}{{Lomb}}{1976}]{1976Ap&SS..39..447L}
{Lomb} N.~R.,  1976, \mn@doi [\apss] {10.1007/BF00648343}, \href
  {http://adsabs.harvard.edu/abs/1976Ap%26SS..39..447L} {39, 447}

\bibitem[\protect\citeauthoryear{{Lucas} et~al.,}{{Lucas}
  et~al.}{2008}]{2008MNRAS.391..136L}
{Lucas} P.~W.,  et~al., 2008, \mn@doi [\mnras]
  {10.1111/j.1365-2966.2008.13924.x}, \href
  {http://adsabs.harvard.edu/abs/2008MNRAS.391..136L} {391, 136}

\bibitem[\protect\citeauthoryear{{Luhman}}{{Luhman}}{2004}]{2004ApJ...602..816L}
{Luhman} K.~L.,  2004, \mn@doi [\apj] {10.1086/381146}, \href
  {http://adsabs.harvard.edu/abs/2004ApJ...602..816L} {602, 816}

\bibitem[\protect\citeauthoryear{{Lynds}}{{Lynds}}{1962}]{1962ApJS....7....1L}
{Lynds} B.~T.,  1962, \mn@doi [\apjs] {10.1086/190072}, \href
  {http://adsabs.harvard.edu/abs/1962ApJS....7....1L} {7, 1}

\bibitem[\protect\citeauthoryear{{Magakian} et~al.,}{{Magakian}
  et~al.}{2013}]{2013MNRAS.432.2685M}
{Magakian} T.~Y.,  et~al., 2013, \mn@doi [\mnras] {10.1093/mnras/stt626}, \href
  {http://adsabs.harvard.edu/abs/2013MNRAS.432.2685M} {432, 2685}

\bibitem[\protect\citeauthoryear{{Makidon}, {Rebull}, {Strom}, {Adams}  \&
  {Patten}}{{Makidon} et~al.}{2004}]{2004AJ....127.2228M}
{Makidon} R.~B.,  {Rebull} L.~M.,  {Strom} S.~E.,  {Adams} M.~T.,   {Patten}
  B.~M.,  2004, \mn@doi [\aj] {10.1086/382237}, \href
  {http://adsabs.harvard.edu/abs/2004AJ....127.2228M} {127, 2228}

\bibitem[\protect\citeauthoryear{{Melikian} \& {Karapetian}}{{Melikian} \&
  {Karapetian}}{2001}]{2001Ap.....44..216M}
{Melikian} N.~D.,  {Karapetian} A.~A.,  2001, \mn@doi [Astrophysics]
  {10.1023/A:1010996726106}, \href
  {http://adsabs.harvard.edu/abs/2001Ap.....44..216M} {44, 216}

\bibitem[\protect\citeauthoryear{{Melikian} \& {Karapetian}}{{Melikian} \&
  {Karapetian}}{2003}]{2003Ap.....46..282M}
{Melikian} N.~D.,  {Karapetian} A.~A.,  2003, \mn@doi [Astrophysics]
  {10.1023/A:1025493412340}, \href
  {http://adsabs.harvard.edu/abs/2003Ap.....46..282M} {46, 282}

\bibitem[\protect\citeauthoryear{{Melikian}, {Karapetian}  \&
  {Gomez}}{{Melikian} et~al.}{2016}]{2016Ap.....59..484M}
{Melikian} N.~D.,  {Karapetian} A.~A.,   {Gomez} J.,  2016, \mn@doi
  [Astrophysics] {10.1007/s10511-016-9451-8}, \href
  {http://adsabs.harvard.edu/abs/2016Ap.....59..484M} {59, 484}

\bibitem[\protect\citeauthoryear{{Meyer}, {Calvet}  \& {Hillenbrand}}{{Meyer}
  et~al.}{1997}]{1997AJ....114..288M}
{Meyer} M.~R.,  {Calvet} N.,   {Hillenbrand} L.~A.,  1997, \mn@doi [\aj]
  {10.1086/118474}, \href {http://adsabs.harvard.edu/abs/1997AJ....114..288M}
  {114, 288}

\bibitem[\protect\citeauthoryear{{Mondal} et~al.,}{{Mondal}
  et~al.}{2010}]{2010AJ....139.2026M}
{Mondal} S.,  et~al., 2010, \mn@doi [\aj] {10.1088/0004-6256/139/5/2026}, \href
  {http://adsabs.harvard.edu/abs/2010AJ....139.2026M} {139, 2026}

\bibitem[\protect\citeauthoryear{{Movsessian}, {Khanzadyan}, {Aspin},
  {Magakian}, {Beck}, {Moiseev}, {Smith}  \& {Nikogossian}}{{Movsessian}
  et~al.}{2006}]{2006A&A...455.1001M}
{Movsessian} T.~A.,  {Khanzadyan} T.,  {Aspin} C.,  {Magakian} T.~Y.,  {Beck}
  T.,  {Moiseev} A.,  {Smith} M.~D.,   {Nikogossian} E.~H.,  2006, \mn@doi
  [\aap] {10.1051/0004-6361:20054609}, \href
  {http://adsabs.harvard.edu/abs/2006A%26A...455.1001M} {455, 1001}

\bibitem[\protect\citeauthoryear{{Nordhagen}, {Herbst}, {Rhode}  \&
  {Williams}}{{Nordhagen} et~al.}{2006}]{2006AJ....132.1555N}
{Nordhagen} S.,  {Herbst} W.,  {Rhode} K.~L.,   {Williams} E.~C.,  2006,
  \mn@doi [\aj] {10.1086/506985}, \href
  {http://adsabs.harvard.edu/abs/2006AJ....132.1555N} {132, 1555}

\bibitem[\protect\citeauthoryear{{Percy}, {Gryc}, {Wong}  \& {Herbst}}{{Percy}
  et~al.}{2006}]{2006PASP..118.1390P}
{Percy} J.~R.,  {Gryc} W.~K.,  {Wong} J.~C.-Y.,   {Herbst} W.,  2006, \mn@doi
  [\pasp] {10.1086/508557}, \href
  {http://adsabs.harvard.edu/abs/2006PASP..118.1390P} {118, 1390}

\bibitem[\protect\citeauthoryear{{Percy}, {Grynko}, {Seneviratne}  \&
  {Herbst}}{{Percy} et~al.}{2010}]{2010PASP..122..753P}
{Percy} J.~R.,  {Grynko} S.,  {Seneviratne} R.,   {Herbst} W.,  2010, \mn@doi
  [\pasp] {10.1086/654826}, \href
  {http://adsabs.harvard.edu/abs/2010PASP..122..753P} {122, 753}

\bibitem[\protect\citeauthoryear{{Planck Collaboration} et~al.,}{{Planck
  Collaboration} et~al.}{2011}]{2011A&A...536A...1P}
{Planck Collaboration} et~al., 2011, \mn@doi [\aap]
  {10.1051/0004-6361/201116464}, \href
  {http://adsabs.harvard.edu/abs/2011A%26A...536A...1P} {536, A1}

\bibitem[\protect\citeauthoryear{{Planck Collaboration} et~al.,}{{Planck
  Collaboration} et~al.}{2016}]{2016A&A...586A.134P}
{Planck Collaboration} et~al., 2016, \mn@doi [\aap]
  {10.1051/0004-6361/201425022}, \href
  {http://adsabs.harvard.edu/abs/2016A%26A...586A.134P} {586, A134}

\bibitem[\protect\citeauthoryear{{Reipurth} \& {Schneider}}{{Reipurth} \&
  {Schneider}}{2008}]{2008hsf1.book...36R}
{Reipurth} B.,  {Schneider} N.,  2008, {Star Formation and Young Clusters in
  Cygnus}.
p.~36

\bibitem[\protect\citeauthoryear{{Rice}, {Wolk}  \& {Aspin}}{{Rice}
  et~al.}{2012}]{2012ApJ...755...65R}
{Rice} T.~S.,  {Wolk} S.~J.,   {Aspin} C.,  2012, \mn@doi [\apj]
  {10.1088/0004-637X/755/1/65}, \href
  {http://adsabs.harvard.edu/abs/2012ApJ...755...65R} {755, 65}

\bibitem[\protect\citeauthoryear{{Scargle}}{{Scargle}}{1982}]{1982ApJ...263..835S}
{Scargle} J.~D.,  1982, \mn@doi [\apj] {10.1086/160554}, \href
  {http://adsabs.harvard.edu/abs/1982ApJ...263..835S} {263, 835}

\bibitem[\protect\citeauthoryear{{Schaefer}}{{Schaefer}}{1983}]{1983ApJ...266L..45S}
{Schaefer} B.~E.,  1983, \mn@doi [\apjl] {10.1086/183975}, \href
  {http://adsabs.harvard.edu/abs/1983ApJ...266L..45S} {266, L45}

\bibitem[\protect\citeauthoryear{{Schmidt}}{{Schmidt}}{1958}]{1958AN....284...76S}
{Schmidt} K.~H.,  1958, \mn@doi [Astronomische Nachrichten]
  {10.1002/asna.19572840209}, \href
  {http://adsabs.harvard.edu/abs/1958AN....284...76S} {284, 76}

\bibitem[\protect\citeauthoryear{{Siess}, {Dufour}  \& {Forestini}}{{Siess}
  et~al.}{2000}]{2000A&A...358..593S}
{Siess} L.,  {Dufour} E.,   {Forestini} M.,  2000, \aap, \href
  {http://adsabs.harvard.edu/abs/2000A%26A...358..593S} {358, 593}

\bibitem[\protect\citeauthoryear{{Stassun}, {Mathieu}, {Mazeh}  \&
  {Vrba}}{{Stassun} et~al.}{1999}]{1999AJ....117.2941S}
{Stassun} K.~G.,  {Mathieu} R.~D.,  {Mazeh} T.,   {Vrba} F.~J.,  1999, \mn@doi
  [\aj] {10.1086/300881}, \href
  {http://adsabs.harvard.edu/abs/1999AJ....117.2941S} {117, 2941}

\bibitem[\protect\citeauthoryear{{Stetson}}{{Stetson}}{1987}]{1987PASP...99..191S}
{Stetson} P.~B.,  1987, \mn@doi [\pasp] {10.1086/131977}, \href
  {http://adsabs.harvard.edu/abs/1987PASP...99..191S} {99, 191}

\bibitem[\protect\citeauthoryear{{Stetson}}{{Stetson}}{1992}]{1992ASPC...25..297S}
{Stetson} P.~B.,  1992, in {Worrall} D.~M.,  {Biemesderfer} C.,   {Barnes} J.,
  eds,  Astronomical Society of the Pacific Conference Series Vol. 25,
  Astronomical Data Analysis Software and Systems I. p.~297

\bibitem[\protect\citeauthoryear{{Wolk}, {Rice}  \& {Aspin}}{{Wolk}
  et~al.}{2013a}]{2013AJ....145..113W}
{Wolk} S.~J.,  {Rice} T.~S.,   {Aspin} C.~A.,  2013a, \mn@doi [\aj]
  {10.1088/0004-6256/145/4/113}, \href
  {http://adsabs.harvard.edu/abs/2013AJ....145..113W} {145, 113}

\bibitem[\protect\citeauthoryear{{Wolk}, {Rice}  \& {Aspin}}{{Wolk}
  et~al.}{2013b}]{2013ApJ...773..145W}
{Wolk} S.~J.,  {Rice} T.~S.,   {Aspin} C.,  2013b, \mn@doi [\apj]
  {10.1088/0004-637X/773/2/145}, \href
  {http://adsabs.harvard.edu/abs/2013ApJ...773..145W} {773, 145}

\makeatother
\end{thebibliography}

\begin{table*}
\renewcommand{\tabcolsep}{2.5pt}
\caption{Catalog of the identified variable stars\tnote{*}.}
\begin{threeparttable}

\label{tab:cat_var}
\begin{tabular}{cccccccccccccc}
\hline \multicolumn{1}{c}{ID} & \multicolumn{1}{c}{RA($J2000$)} & \multicolumn{1}{c}{Dec($J2000$)}  & \multicolumn{1}{c}{$V-I$}& \multicolumn{1}{c}{$R-I$} & \multicolumn{1}{c}{$I$} & \multicolumn{1}{c}{$J$} & \multicolumn{1}{c}{$H$} & \multicolumn{1}{c}{$K$} & \multicolumn{1}{c}{W1} & \multicolumn{1}{c}{W2}& \multicolumn{1}{c}{W3} & \multicolumn{1}{c}{W4}\\ 

\multicolumn{1}{c}{} & \multicolumn{1}{c}{(hh:mm:ss)} & \multicolumn{1}{c}{(dd:mm:ss)} & \multicolumn{1}{c}{(mag)} & \multicolumn{1}{c}{(mag)}& \multicolumn{1}{c}{(mag)} & \multicolumn{1}{c}{(mag)} & \multicolumn{1}{c}{(mag)}& \multicolumn{1}{c}{(mag)} & \multicolumn{1}{c}{(mag)} & \multicolumn{1}{c}{(mag)}& \multicolumn{1}{c}{(mag)} & \multicolumn{1}{c}{(mag)}\\ \hline
    8 &  20:57:59.76  &  +52:13:01.74  &  $...$ &  1.091 & 19.887 & 16.543 & 15.830 & 15.480 & $...$ &  $...$ &  $...$ &  $...$ \\
  &    &    &    & $\pm$ 0.074 & $\pm$ 0.036 & $\pm$ 0.010 & $\pm$ 0.009 & $\pm$ 0.015 &   &   &   &    \\
   70 &  20:58:1.56  &  +52:20:9.53  &   2.140 &  0.693 & 15.486 & 13.566 & 13.168 & 12.960 & 12.742 &  12.705 &  12.745 &   9.293 \\
  &    &    &  $\pm$ 0.167 & $\pm$ 0.030 & $\pm$ 0.022 & $\pm$ 0.001 & $\pm$ 0.001 & $\pm$ 0.002 & $\pm$ 0.036 & $\pm$ 0.037 & $...$  & $...$   \\
   97 &  20:58:2.64  &  +52:11:59.93  &   2.289 &  0.846 & 18.464 & 16.235 & 15.637 & 15.419 & 14.156 &  14.348 &  12.811 &   9.229 \\
  &    &    &  $\pm$ 0.105 & $\pm$ 0.077 & $\pm$ 0.061 & $\pm$ 0.008 & $\pm$ 0.008 & $\pm$ 0.014 & $\pm$ 0.025 & $\pm$ 0.037 & $\pm$ 0.456 & $\pm$ 0.425  \\
  104 &  20:58:2.64  &  +52:21:20.81  &   1.798 &  0.772 & 16.127 & 14.049 & 13.527 & 13.317 & 11.921 &  11.930 &  11.390 &   9.168 \\
  &    &    &  $\pm$ 0.045 & $\pm$ 0.031 & $\pm$ 0.025 & $\pm$ 0.002 & $\pm$ 0.001 & $\pm$ 0.003 & $\pm$ 0.022 & $\pm$ 0.022 & $\pm$ 0.132 & $\pm$ 0.426  \\
  567 &  20:58:16.68  &  +52:11:11.98  &  $...$ & $...$ & 16.331 & 13.065 & 11.937 & 11.582 & 11.268 &  11.318 &  11.500 &   9.347 \\
  &    &    &    &   & $\pm$ 0.040 & $\pm$ 0.027 & $\pm$ 0.027 & $\pm$ 0.020 & $\pm$ 0.023 & $\pm$ 0.021 & $\pm$ 0.141 &  $...$  \\
  750 &  20:58:22.08  &  +52:12:12.10  &  $...$ &  1.063 & 18.192 & 15.120 & 14.492 & 14.171 & 13.391 &  13.423 &  11.548 &   9.191 \\
  &    &    &    & $\pm$ 0.055 & $\pm$ 0.033 & $\pm$ 0.003 & $\pm$ 0.003 & $\pm$ 0.005 & $\pm$ 0.028 & $\pm$ 0.031 & $\pm$ 0.161 & $\pm$ 0.497  \\
  771 &  20:58:22.44  &  +52:16:25.61  &  $...$ &  0.874 & 20.396 & 17.071 & 16.362 & 16.044 & $...$ &  $...$ &  $...$ &  $...$ \\
  &    &    &    & $\pm$ 0.070 & $\pm$ 0.062 & $\pm$ 0.016 & $\pm$ 0.014 & $\pm$ 0.025 &   &   &   &    \\
  832 &  20:58:24.24  &  +52:11:52.04  &   1.663 &  0.740 & 15.877 & 13.843 & 13.422 & 13.193 & 12.447 &  12.462 &  12.377 &   9.181 \\
  &    &    &  $\pm$ 0.034 & $\pm$ 0.034 & $\pm$ 0.020 & $\pm$ 0.001 & $\pm$ 0.001 & $\pm$ 0.002 & $\pm$ 0.026 & $\pm$ 0.027 & $\pm$ 0.423 &  $...$  \\
  954 &  20:58:27.12  &  +52:16:34.43  &  $...$ &  1.237 & 21.399 & 17.992 & 17.358 & 17.091 & $...$ &  $...$ &  $...$ &  $...$ \\
  &    &    &    & $\pm$ 0.117 & $\pm$ 0.080 & $\pm$ 0.035 & $\pm$ 0.035 & $\pm$ 0.064 &   &   &   &    \\
 1142 &  20:58:33.96  &  +52:15:41.04  &   1.779 &  0.807 & 16.593 & 14.210 & 13.721 & 13.495 & 12.545 &  12.632 &  12.322 &   9.350 \\
  &    &    &  $\pm$ 0.043 & $\pm$ 0.010 & $\pm$ 0.006 & $\pm$ 0.002 & $\pm$ 0.002 & $\pm$ 0.003 & $\pm$ 0.024 & $\pm$ 0.025 &  $...$ &  $...$  \\
 1170 &  20:58:34.68  &  +52:12:20.77  &  $...$ &  1.396 & 19.677 & 16.658 & 15.910 & 15.573 & $...$ &  $...$ &  $...$ &  $...$ \\
  &    &    &    & $\pm$ 0.063 & $\pm$ 0.056 & $\pm$ 0.011 & $\pm$ 0.010 & $\pm$ 0.016 &   &   &   &    \\
 1259 &  20:58:38.28  &  +52:11:50.21  &   1.805 &  0.925 & 18.050 & 15.207 & 14.499 & 14.217 & 13.780 &  13.740 &  12.748 &   9.371 \\
  &    &    &  $\pm$ 0.117 & $\pm$ 0.024 & $\pm$ 0.011 & $\pm$ 0.004 & $\pm$ 0.003 & $\pm$ 0.005 & $\pm$ 0.026 & $\pm$ 0.036 & $...$  & $...$   \\
 1269 &  20:58:38.64  &  +52:14:26.30  &   2.379 &  1.071 & 18.394 & 15.483 & 14.842 & 14.497 & 13.104 &  13.090 &  12.661 &   9.407 \\
  &    &    &  $\pm$ 0.086 & $\pm$ 0.056 & $\pm$ 0.044 & $\pm$ 0.004 & $\pm$ 0.004 & $\pm$ 0.007 & $\pm$ 0.026 & $\pm$ 0.030 &  $...$ &   $...$ \\
 1422 &  20:58:43.68  &  +52:13:16.82  &  $...$ &  1.122 & 20.288 & 17.143 & 16.324 & 16.062 & $...$ &  $...$ &  $...$ &  $...$ \\
  &    &    &    & $\pm$ 0.116 & $\pm$ 0.068 & $\pm$ 0.017 & $\pm$ 0.014 & $\pm$ 0.026 &   &   &   &    \\
 1509 &  20:58:46.92  &  +52:19:41.88  &  $...$ &  0.755 & 19.793 & 17.127 & 16.287 & 15.976 & $...$ &  $...$ &  $...$ &  $...$ \\
  &    &    &    & $\pm$ 0.090 & $\pm$ 0.053 & $\pm$ 0.016 & $\pm$ 0.013 & $\pm$ 0.023 &   &   &   &    \\
 1516 &  20:58:47.64  &  +52:19:4.84  &  $...$ & $...$ & 20.504 & 17.361 & 16.468 & 16.144 & $...$ &  $...$ &  $...$ &  $...$ \\
  &    &    &    &   & $\pm$ 0.071 & $\pm$ 0.020 & $\pm$ 0.016 & $\pm$ 0.027 &   &   &   &    \\
 1549 &  20:58:49.08  &  +52:12:29.41  &  $...$ & $...$ & 20.178 & 15.795 & 14.590 & 14.071 & $...$ &  $...$ &  $...$ &  $...$ \\
  &    &    &    &   & $\pm$ 0.046 & $\pm$ 0.005 & $\pm$ 0.003 & $\pm$ 0.005 &   &   &   &    \\
 1564 &  20:58:49.80  &  +52:19:46.60  &   2.117 &  0.771 & 15.461 & 12.784 & 11.770 & 11.113 & 10.259 &   9.839 &   8.269 &   6.307 \\
  &    &    &  $\pm$ 0.118 & $\pm$ 0.026 & $\pm$ 0.011 & $\pm$ 0.026 & $\pm$ 0.023 & $\pm$ 0.024 & $\pm$ 0.022 & $\pm$ 0.020 & $\pm$ 0.023 & $\pm$ 0.049  \\
 1633 &  20:58:52.32  &  +52:18:0.22  &  $...$ &  1.092 & 20.259 & 17.348 & 16.467 & 16.136 & 14.563 &  14.765 &  12.472 &   8.825 \\
  &    &    &    & $\pm$ 0.135 & $\pm$ 0.089 & $\pm$ 0.020 & $\pm$ 0.016 & $\pm$ 0.029 & $\pm$ 0.032 & $\pm$ 0.051 & $\pm$ 0.358 &  $...$  \\
 1648 &  20:58:53.04  &  +52:18:44.57  &   1.042 &  0.482 & 13.937 & 12.556 & 12.097 & 11.966 & 11.926 &  12.035 &  12.024 &   9.155 \\
  &    &    &  $\pm$ 0.017 & $\pm$ 0.010 & $\pm$ 0.006 & $\pm$ 0.017 & $\pm$ 0.017 & $\pm$ 0.027 & $\pm$ 0.022 & $\pm$ 0.021 & $\pm$ 0.286 &  $...$  \\
 1751 &  20:58:59.16  &  +52:18:16.78  &   3.677 &  1.581 & 17.310 & 12.273 & 10.972 & 10.486 & 10.122 &  10.141 &   9.904 &   9.401 \\
  &    &    &  $\pm$ 0.062 & $\pm$ 0.012 & $\pm$ 0.006 & $\pm$ 0.020 & $\pm$ 0.017 & $\pm$ 0.022 & $\pm$ 0.023 & $\pm$ 0.021 & $\pm$ 0.043 &  $...$  \\
 1771 &  20:58:59.88  &  +52:22:18.48  &   2.283 &  1.009 & 15.722 & 12.425 & 11.248 & 10.577 & 10.193 &   9.615 &   7.288 &   4.832 \\
  &    &    &  $\pm$ 0.039 & $\pm$ 0.032 & $\pm$ 0.013 & $\pm$ 0.019 & $\pm$ 0.019 & $\pm$ 0.020 & $\pm$ 0.023 & $\pm$ 0.021 & $\pm$ 0.021 & $\pm$ 0.029  \\
 1786 &  20:59:0.24  &  +52:13:6.53  &   3.096 &  1.362 & 16.564 & 12.357 & 11.294 & 10.892 & 10.637 &  10.669 &  10.629 &   8.856 \\
  &    &    &  $\pm$ 0.026 & $\pm$ 0.009 & $\pm$ 0.005 & $\pm$ 0.022 & $\pm$ 0.024 & $\pm$ 0.027 & $\pm$ 0.028 & $\pm$ 0.027 & $\pm$ 0.095 & $...$   \\
 3451 &  20:57:47.52  &  +52:20:35.77  &  $...$ & $...$ & 18.200 & 15.499 & 14.965 & 14.730 & 14.597 &  14.771 &  13.127 &   9.405 \\
  &    &    &    &   & $\pm$ 0.080 & $\pm$ 0.004 & $\pm$ 0.004 & $\pm$ 0.008 & $\pm$ 0.032 & $\pm$ 0.054 &  $...$ &  $...$  \\
 3985 &  20:57:58.32  &  +52:8:13.70  &  $...$ & $...$ & 17.069 & 14.022 & 13.383 & 13.165 & 12.767 &  12.819 &  12.635 &   9.262 \\
  &    &    &    &   & $\pm$ 0.089 & $\pm$ 0.002 & $\pm$ 0.001 & $\pm$ 0.002 & $\pm$ 0.024 & $\pm$ 0.025 &  $...$ &  $...$  \\
 6087 &  20:58:18.48  &  +52:10:18.34  &  $...$ & $...$ & 20.367 & 17.224 & 16.449 & 16.084 & $...$ &  $...$ &  $...$ &  $...$ \\
  &    &    &    &   & $\pm$ 0.076 & $\pm$ 0.018 & $\pm$ 0.016 & $\pm$ 0.025 &   &   &   &    \\
 6289 &  20:59:11.04  &  +52:11:51.00  &   1.830 &  0.868 & 16.030 & 13.724 & 13.269 & 13.053 & 10.829 &  10.830 &  11.088 &   9.355 \\
  &    &    &  $\pm$ 0.022 & $\pm$ 0.020 & $\pm$ 0.019 & $\pm$ 0.001 & $\pm$ 0.001 & $\pm$ 0.002 & $\pm$ 0.023 & $\pm$ 0.021 & $\pm$ 0.116 &  $...$  \\
 6506 &  20:58:55.20  &  +52:13:34.75  &  $...$ &  1.304 & 19.935 & 16.363 & 15.563 & 15.163 & 13.429 &  13.473 &  11.947 &   9.001 \\
  &    &    &    & $\pm$ 0.098 & $\pm$ 0.051 & $\pm$ 0.009 & $\pm$ 0.007 & $\pm$ 0.012 & $\pm$ 0.024 & $\pm$ 0.029 & $\pm$ 0.193 & $...$   \\
 6666 &  20:59:10.68  &  +52:15:20.10  &   2.004 &  0.989 & 14.439 & 11.558 & 11.015 & 10.741 & 10.591 &  10.519 &  10.637 &   8.828 \\
  &    &    &  $\pm$ 0.021 & $\pm$ 0.006 & $\pm$ 0.006 & $\pm$ 0.015 & $\pm$ 0.015 & $\pm$ 0.022 & $\pm$ 0.022 & $\pm$ 0.021 & $\pm$ 0.077 &  $...$  \\
 6747 &  20:59:2.76  &  +52:16:27.98  &   3.242 &  1.442 & 16.077 & 11.715 & 10.871 & 10.446 & 10.371 &  10.136 &  10.347 &   9.434 \\
  &    &    &  $\pm$ 0.054 & $\pm$ 0.007 & $\pm$ 0.004 & $\pm$ 0.015 & $\pm$ 0.015 & $\pm$ 0.022 & $\pm$ 0.022 & $\pm$ 0.020 & $\pm$ 0.061 & $\pm$ 0.532  \\
 6927 &  20:59:0.60  &  +52:18:41.22  &   3.156 &  1.390 & 16.760 & 12.388 & 11.329 & 10.912 & 10.514 &  10.514 &  10.319 &   8.711 \\
  &    &    &  $\pm$ 0.031 & $\pm$ 0.012 & $\pm$ 0.008 & $\pm$ 0.023 & $\pm$ 0.026 & $\pm$ 0.032 & $\pm$ 0.022 & $\pm$ 0.020 & $\pm$ 0.066 &  $...$  \\

\hline\end{tabular}
\begin{tablenotes}\footnotesize
 \item [*] The optical and NIR magnitudes with error $<$ 0.1 mag and WISE magnitudes with error $<$ 0.2 mag were used for this analyses. The magnitudes with null error information were not considered in this study.
\end{tablenotes}
\end{threeparttable}
\end{table*}

\begin{table*}

\centering
\renewcommand{\tabcolsep}{2.5pt}
\caption{Details of the variable stars.}
\begin{threeparttable}
\label{tab:parameters}
\begin{tabular}{ccccccccccccc}
\hline \multicolumn{1}{c}{ID} & \multicolumn{1}{c}{$I_{mean}$} & \multicolumn{1}{c}{RMS}  & \multicolumn{1}{c}{Period}& \multicolumn{1}{c}{FAP} & \multicolumn{1}{c}{Amp} & \multicolumn{1}{c}{Spectral Index} & \multicolumn{1}{c}{Chisq}  & \multicolumn{1}{c}{Age\tnote{*}} & \multicolumn{1}{c}{Mass\tnote{*}} & \multicolumn{1}{c}{Remarks}\\ 

 \multicolumn{1}{c}{} & \multicolumn{1}{c}{(mag)} & \multicolumn{1}{c}{(mag)}  & \multicolumn{1}{c}{(days)}& \multicolumn{1}{c}{}& \multicolumn{1}{c}{(mag)} & \multicolumn{1}{c}{($\alpha$)}  & \multicolumn{1}{c}{}  & \multicolumn{1}{c}{(Myr)} & \multicolumn{1}{c}{($M\sun$)} & \multicolumn{1}{c}{}\\ 
\hline 
   8 &  19.89 & 0.186 &  6.67 & $<$ 0.01 &0.41   &   $...$ &    $...$      &  $...$ & $...$  & $\delta$-scuti \\
  70 &  15.49 & 0.044 &  0.28 &  $<$ 0.01 & 0.05   &  $...$ &  $...$   & $...$  & $...$ & EB \\
  97 &  18.46 & 0.111 &   $...$ & $...$&  $...$   &  $...$ & $...$   & $...$  & $...$ & $...$\\
 104 &  16.13 & 0.050 &  $...$ &   $...$ &  $...$   & -1.92  & 0.92   & $...$  & $...$ & MS/Field\\
 567 &  16.33 & 0.125 &  $...$&   $...$  &  $...$   &  -2.84 & 0.99  & $...$  & $...$ & MS/Field \\
 750 &  18.19 & 0.10 &  $...$ &   $...$ &  $...$   &  -1.41  & 0.95   & $...$  & $...$ & CTTS \\
 771 &  20.40 & 0.18 &  $...$&   $...$  &  $...$   &  $...$   &  $...$   & $...$  & $...$ & $...$\\
 832 &  15.88 & 0.07 &  $...$ & $...$& $...$   &  $...$& $...$  & $...$  & $...$ & $...$ \\
 954 &  21.40 & 0.23 &  0.43 &  $<$ 0.01  &  0.27   &  $...$ &  $...$ & $...$  & $...$& $...$ \\
1142 &  16.59 & 0.07 &  $...$   &  $...$    &  $...$ & $...$    & $...$  & $...$& $...$\\
1170 &  19.68 & 0.14 &  11.58 &  $<$ 0.01 &  0.35   &  $...$ &  $...$ &  $...$ & $...$  & $\delta$-scuti \\ 
1259 &  18.05 & 0.09 &  0.30&  $<$ 0.01  &  0.20   &  $...$ & $...$    & $...$  & $...$&  Ex-Lupi\\
1269 &  18.39 & 0.19 &  2.40 &  $<$ 0.01 &  0.16   &  $...$ & $...$   & $...$  & $...$& $...$ \\
1422 &  20.29 & 0.15 &  $...$ & $...$ & $...$   &  $...$  & $...$ & $...$ & $...$  & $...$\\
1509 &  19.79 & 0.17 &  $...$  &$...$ &  $...$    & $...$  & $...$   & $...$  & $...$  & $...$\\
1516 &  20.50 & 0.26 &  0.58 &  $<$ 0.01 &  0.62   & $...$ & $...$  & $...$ & $...$  & $...$ \\
1549 &  20.18 & 0.16 &  $...$ & $...$ & $...$   & $...$ & $...$  & $...$  & $...$ & $...$\\
1564 &  15.46 & 0.11 &  5.17 &  $<$ 0.01  &  0.31   &  -1.01 & 0.94  &  5.12$\pm$0.88 &  0.80$\pm$0.03& CTTS  \\
1633 &  20.26 & 0.15 &  9.87 &  $<$ 0.01  &  0.19   &   $...$ & $...$  & $...$  & $...$& $...$ \\
1648 &  13.94 & 0.05 &  $...$ &  $...$   &  $...$ &  $...$ & $...$  & $...$& $...$&   $...$\\
1751 &  17.31 & 0.07 &  $...$&  $...$ &  $...$   &   -2.55 & 0.99  &  1.82$\pm$0.51 &  0.13$\pm$ 0.02& MS/Field\\
1771 &  15.72 & 0.16 &  $...$  &  $...$  &   $...$  &  -0.31  & 0.90  &  4.46$\pm$ 1.12 &  0.67$\pm$0.04& CTTS\\
1786 &  16.56 & 0.06 &  $...$ &  $...$ &   $...$  &  -2.73 & 0.99  & 3.97$\pm$0.76 &  0.24$\pm$0.02& MS/Field \\
3451 &  18.20 & 0.10 &  $...$  &  $...$  &   $...$  &  $...$  &  $...$  & $...$  & $...$& $...$\\
3985 &  17.07 & 0.05 &  0.15&  $<$ 0.01  &  0.15   &  -2.40  &  0.99 & $...$ & $...$  & MS/Field; EB\\
6087 &  20.37 & 0.16 &  $...$ &   $...$ & $...$   &  $...$ &  $...$  & $...$  & $...$& $...$ \\
6289 &  16.03 & 0.11 &  1.70 &  $<$ 0.01 &  0.18   &  -2.03&  0.75   & $...$  & $...$& MS/Field; EB\\
6506 &  19.93 & 0.25 &  1.18&  $<$ 0.01  &  0.72   &  -1.21 &  0.80 & $...$  & $...$& CTTS\\
6666 &  14.44 & 0.07 &  2.44 &  $<$ 0.01 &  0.09   &  -2.80 &  0.99  &  2.22$\pm$0.29 &  0.70$\pm$0.04& MS/Field; EB\\
6747 &  16.08 & 0.09 &  1.58&  $<$ 0.01  &  0.09   &  -2.79 &  0.99    &  2.25$\pm$0.75 &  0.23$\pm$0.02& MS/Field; EB \\
6927 &  16.76 & 0.07 &  $...$ &  $...$ &  $...$   &  -2.33 &  0.99  &  3.03$\pm$ 0.34 &  0.20$\pm$0.03& MS/Field \\

\hline\end{tabular}
\begin{tablenotes}\footnotesize
 \item [*] Age and Mass are estimated from $V/V-I$ CMD.
\end{tablenotes}
\end{threeparttable}
\end{table*}





\end{document}